\documentclass[twocolumn,twocolappendix]{aastex701}

\usepackage{nccmath}
\usepackage{mathtools}
\usepackage{placeins}

\begin{document}

\title{Predicting Multiwavelength Emission Associated with X-Ray Flares and Extended Emission of Gamma-Ray Bursts}
\correspondingauthor{Riki Matsui}

\author[0000-0003-0805-7741]{Riki Matsui}
\affiliation{Astronomical Institute, Graduate School of Science, Tohoku University, Sendai 980-8578, Japan}
\affiliation{Department of Earth Science and Astronomy, Graduate School of Arts and Sciences, The University of Tokyo, Meguro, Tokyo 153-8902, Japan}
\email[show]{riki.matsui@astr.tohoku.ac.jp}  

\author[0000-0003-2579-7266]{Shigeo S. Kimura}
\affiliation{Frontier Research Institute for Interdisciplinary Sciences, Tohoku University, Sendai 980-8578, Japan}
\affiliation{Astronomical Institute, Graduate School of Science, Tohoku University, Sendai 980-8578, Japan}
\email{shigeo@astr.tohoku.ac.jp}  

\author[0000-0002-5358-5642]{Kohta Murase}
\affiliation{Department of Astronomy and Astrophysics, The Pennsylvania State University, 525 Davey Laboratory, University Park, PA 16802, USA}
\affiliation{Institute for Gravitation and the Cosmos, The Pennsylvania State University, University Park, PA 16802, USA}
\affiliation{Department of Physics; Department of Astronomy \& Astrophysics; Center for Multimessenger Astrophysics, Institute for Gravitation and the Cosmos, The Pennsylvania State University, University Park, PA 16802, USA}
\affiliation{Center for Gravitational Physics and Quantum Information,
Yukawa Institute for Theoretical Physics, Kyoto University, Kyoto, Kyoto 606-8502, Japan}
\email{murase@psu.edu}

\author[0000-0003-2478-333X]{B. Theodore Zhang}
\affiliation{Key Laboratory of Particle Astrophysics and Experimental Physics Division and Computing Center, Institute of High Energy Physics, Chinese Academy of Sciences, 100049 Beijing, China}
\affiliation{TIANFU Cosmic Ray Research Center, Chengdu, Sichuan, China}
\email{zhangbing@ihep.ac.cn}



\begin{abstract}
Gamma-ray bursts (GRBs) are one of the most extreme transients in the universe, but their explosion and emission mechanism remains unclear. 
To investigate the nature of GRB jets, here we focus on X-ray flares (XFs) and extended emissions (EEs), which are X-ray emissions that occur 100 to 1000 seconds after the main burst. They can be observed by recently developed multi-wavelength facilities. In this paper, we calculate emissions across multi-wavelengths associated with XFs and EEs under the hypothesis that XFs and EEs are optically-thin synchrotron emissions from nonthermal electrons in relativistic jets. Considering ranges of the dissipation radius $r_{\rm diss}$ and the Lorentz factor $\Gamma$ of the jet, we determine the parameter space in which a detectable emission can be produced at each wavelength.
We found that simultaneous ultraviolet and very-high-energy gamma-ray emission associated with XFs or EEs can be detected by Swift/UVOT, SVOM/VT, and CTAO approximately every three years. The detection and non-detection rates for each detector are key to determining the uncertain yet essential values necessary for understanding the physics of GRB jets.
\end{abstract}



\section{Introduction} 
\label{sec:intro}
Gamma-ray bursts (GRBs) are one of the most extreme transients in the universe.
Their isotropic-equivalent gamma-ray luminosity typically reaches $10^{51}$ erg/s and such luminous emission is powered by a relativistic jet \citep{Schmidt1978Natur.271..525S,Paczynski1986,Goodman1986ApJ...308L..47G}.
The nonthermal spectrum of the burst indicates that nonthermal dissipation within the jet may produce the radiation \citep{Band1993ApJ...413..281B}.
However, the origin of their jets and gamma-ray emissions remains unresolved.

The main gamma-ray emission of GRBs, the so-called prompt emission, typically lasts around 10~s \citep{Kouveliotou1993}.
Their prompt emissions require around ten-seconds reaction time for facilities, which makes difficult to observe at multi-wavelength.
This prevents comprehensive multi-wavelength observations and a deeper understanding of the jet, except for a few special events involving prompt-like detections in the optical and GeV bands, such as GRB 080319B \citep{Racusin2008NakedEye}, GRB 220101A \citep{Jin2023_GRB220101}, and GRB 221009A \citep{Axelsson-Fermi}.

On the other hand, X-ray emission with $\sim 10^{48}$-$10^{49}$ erg/s following the prompt phase lasts about 1000 s and appears as either an X-ray flare (XF) or extended emission (EE) \citep{Piro2005ApJ...623..314P,Burrows2005Sci...309.1833B,Nousek2006ApJ...642..389N,Norris2006ApJ...643..266N}.
These types of X-ray emission are observed in about one-third to one-half of all GRBs \citep{Sakamoto2011ApJS..195....2S,Swenson2014-XF,Kaneko2015MNRAS.452..824K,Kagawa2015ApJ...811....4K,Kagawa2019ApJ...877..147K,Lien2016ApJ...829....7L,Liu2019ApJ...884...59L}.
They are roughly 100-1000 times less luminous than the prompt emission, but their integrated energy is comparable \citep{Yi2016ApJS..224...20Y,Kisaka2017ApJ...846..142K}.
Moreover, the short-timescale variability in their light curves is a common feature with the prompt emission \citep{Swenson2013UVF,Swenson2014-XF}. 
Although the delay in these emissions indicates that they likely arise from another late jet component \footnote{The slightly off-axis prompt emission may explain some part of the flares \citep{Duque2022offaxisFlare}.} \citep{Ioka2005ApJ...631..429I,Perna2006ApJ...636L..29P,Zhang2006ApJ...642..354Z,Gao2006ChJAA...6..513G,Metzger2008MNRAS.385.1455M,Rowlinson2013MNRAS.430.1061R,Gompertz2014MNRAS.438..240G,Kisaka2015ApJ...804L..16K}, the dissipation process could be internal dissipation of jets as in the prompt jets \citep{Fan2005MNRAS.364L..42F,Ioka2005ApJ...631..429I,Falcone2006ApJ...641.1010F,Liang2006ApJ...646..351L,Zhang2006ApJ...642..354Z}. 
The late-time jets that emit XFs and EEs and its dissipation mechanism could be completely different from that of the prompt emission. However, understanding their dissipation should be an important first step in realizing the dissipation process of GRB jets.

In addition to XFs and EEs, ultraviolet (UV) flares are detected with a comparable frequency \citep{Swenson2013UVF,Yi2017_UVF}.
Although it is unclear whether UV flares are correlated with XFs or EEs, there are a few flares clearly associated with XFs \citep{Yi2017_UVF,Becerra2021_GRB1803325}.
These events suggest the possibility of multi-wavelength emission lasting about 1000 s after the main burst of GRBs.

Here, we calculate the broadband emission associated with XFs and EEs, hypothesizing that they are produced by synchrotron radiation from nonthermal electrons in the relativistic jet.
Based on this hypothesis, we consistently solve the transport equation while accounting for the full leptonic processes, such as synchrotron emission and self-absorption, inverse Compton scattering, and two-photon ($\gamma\gamma$) absorption.
By considering a wide range of dissipation radii $r_{\rm diss}$ and jet Lorentz factors $\Gamma$\footnote{Previous studies have constrained the bulk Lorentz factor of jets that produce XFs and EEs \citep{Falcone2006ApJ...641.1010F,Yi2015ApJ...807...92Y,Matsumoto2020MNRAS.493..783M}, but these constraints are based on the model that each study employs.
Here, we seek independent constraints on this parameter.}, we clarify where in the parameter space spectral features of absorption processes can or cannot be detected.
Our predictions can be used to constrain the values of $r_{\rm diss}$ and $\Gamma$ in GRB jets based on future observations. 
This will lead to a better understanding of the physics of the jets \citep{Rees1994ApJ...430L..93R,Rees2005ApJ...628..847R,Zhang2011ApJ...726...90Z,Zhang2013PhRvL.110l1101Z,Troja2015ApJ...803...10T,Yi2015ApJ...807...92Y,Matsumoto2020MNRAS.493..783M}.

This paper is organized as follows.
Section \ref{sec:method} presents the method and the setup of our calculations.
Section \ref{sec:result} and \ref{sec:detection} provide the results and the observational expectations, respectively.
Section \ref{sec:discussion} discusses a possible contaminated component, an application, and extensions of this work.
Finally, the conclusion is given in Section \ref{sec:conclusion}.

Throughout this paper, we use the notation $Q_X = Q/10^X$ in cgs units unless otherwise noted, and denote by $Q'$ the physical quantities in the comoving frame of the jet.
We adopt $H_0 = 70\ \rm km \ s^{-1}\ Mpc^{-1}$, $\Omega_{\rm M} = 0.30$, and $\Omega_{\Lambda} = 0.70$ as cosmological parameters.
In addition, in this paper, $z$ and $d_L$ are used as the redshift and the luminosity distance, respectively.
 
We define the commonly used physical constants as follows:
$c$ is the speed of light, $\sigma_T$ is the Thomson scattering cross section, $m_e$ is the electron mass, $m_p$ is the proton mass, $h$ is the Planck constant, $e$ is the elementary charge.

\section{method}
\label{sec:method}
Here, we calculate leptonic synchrotron emission for XFs and EEs.
Nonthermal synchrotron emission with high radiation efficiency cannot explain the typical broken power-law spectra of prompt emission, especially in the low-energy part \citep{Zhang2011ApJ...726...90Z}.
However, it could be applicable to XFs and EEs, since they are well fitted by a single power law \citep{Falcone2007_XFspectra,Sakamoto2011ApJS..195....2S,Lien2016ApJ...829....7L,Liu2019_XFsubclass}.
Even though some studies for luminous XFs exhibit a variety of spectral shapes \citep{Falcone2007_XFspectra,Peng2014XF_spectra,Uhm2016XF,Geng2017XF,Liu2019_XFsubclass,Ajello2019_131108A,Zhou2025_170519}\footnote{
Some resemble prompt emission, others contain a thermal component, and still others show a very hard single power-law with exponential cutoff.}, the spectral break is unclear for the less luminous, mainly populated XFs and EEs \citep{Falcone2007_XFspectra}\footnote{Low mass accretion rate $\sim 10^{-3}~M_\odot$/s, required by low-luminosity jets, cannot ignite the neutrino production \citep{Lei2017NDAF}. The different state of the jet base and the jet composition may involve different dissipation mechanisms and spectral features from those of high-luminosity jets. In this work, we focus on XFs with X-ray luminosities of $\sim 10^{49}$ erg/s and assume an electron injection with $p = 2$ for simplicity.}.
The photon index of the single power-low fitting for typical XFs and EEs is $\sim1.5$-$2$, which is consistent with the synchrotron emission from nonthermal electrons injected with single power-law index $p\sim 2$ \citep{Dermer09book} and efficiently cooled by the process.
Even if XFs and EEs exhibit broken power-law spectra like prompt emissions, our qualitative picture may still apply.

To estimate the emission spectrum, we adopt a one-zone approximation.
The dissipation region is fixed at a radius $r_{\rm diss}$, where a jet with Lorentz factor $\Gamma$ dissipates its isotropic luminosity $L_{\rm e,iso}$ into electrons. 
We follow a representative fluid element advecting from $r_{\rm diss}$ to $2r_{\rm diss}$, assuming that its physical quantities represent spatially, directionally, and temporally averaged values within the dissipation region.
The momentum-differential number density of electrons $n^\prime_{p^\prime_e}$ and the energy-differential photon number density $n^\prime_{\varepsilon^\prime_\gamma}$ are assumed to be uniform and isotropic.
The magnetic field at the dissipation region is also assumed to be uniform.

The following provides a qualitative picture of the radiation process on the comoving frame of the fluid element.
The dissipated electrons are continuously injected at a rate $\dot n^\prime_{p^\prime_e,\rm diss}$ and are cooled by adiabatic expansion on a timescale $t^\prime_{\rm dyn} = r_{\rm diss}/(c\Gamma)$.
The electrons emit synchrotron radiation with an energy-differential and angle-integrated emissivity $j^\prime_{\varepsilon^\prime_\gamma,\rm syn}$ and are cooled on a timescale $t_{\rm syn}^\prime$.
A fraction of the radiation is absorbed through synchrotron self-absorption (SSA) on a timescale $t_{\rm SSA}^\prime$ or through $\gamma\gamma$ absorption on a timescale $t_{\rm \gamma\gamma}^\prime$.
Electrons and positrons are produced by $\gamma\gamma$ interactions at a rate $\dot n^\prime_{p^\prime_e,\gamma\gamma}$, and they emit synchrotron radiation again.
In addition, Compton scattering redistributes both electrons and photons.
We define $t^\prime_{\rm ic}$ as the inverse Compton cooling timescale for electrons, $t^\prime_{\gamma,\rm ic}$ as the Compton scattering timescale for photons, and $j^\prime_{\varepsilon^\prime_\gamma,\rm ic}$ as the emissivity of inverse Compton scattering.
The photons produced through these processes escape from the emission region on a timescale $t^\prime_{\rm esc}$.
This timescale is defined as the light crossing time of the radial width of the dissipation region $\Delta r^\prime$, therefore, $t^\prime_{\rm esc} = \Delta r^\prime / c$.
Using this $t^\prime_{\rm esc}$, The observed flux is proportional to $n^\prime_{\varepsilon^\prime_\gamma}/t^\prime_{\rm esc}$.

We describe the above processes using the transport equations in momentum space \citep{Dermer09book}:
\begin{equation}
\label{eq:transport}
\begin{split}
&\frac{\partial n^\prime_{p^\prime_e}}{\partial t^\prime} - \frac{\partial }{\partial p^\prime_e }\left(\frac{p^\prime_e n^\prime_{p^\prime_e}}{t^\prime_{\rm cool}}\right) = \dot n^\prime_{p^\prime_e,\rm diss}+\dot n^\prime_{p^\prime_e,\gamma\gamma},\\
&\frac{\partial n^\prime_{\varepsilon^\prime_\gamma}}{\partial t^\prime} = -\frac{ n^\prime_{\varepsilon^\prime_\gamma}}{t^\prime_{\rm dyn}}-\frac{ n^\prime_{\varepsilon^\prime_\gamma}}{t^\prime_{\rm abs}}+j^\prime_{ \varepsilon^\prime_\gamma},
\end{split}
\end{equation}
where $t^\prime$ is the time, $p^\prime_e$ is the momentum of an electron or positron, $\varepsilon_\gamma^\prime$ is the photon energy, $t_{\rm cool}^\prime = (t_{\rm syn}^{\prime-1}+t_{\rm ic}^{\prime-1}+t_{\rm dyn}^{\prime-1})^{-1}$ is the total cooling timescale, $t_{\rm abs}^\prime= (t_{\rm \gamma\gamma}^{\prime-1}+t_{\rm SSA}^{\prime-1}+t_{\rm \gamma,ic}^{\prime-1})^{-1}$ is the total effective absorption timescale, and $j^\prime_{ \varepsilon^\prime_\gamma} = j^\prime_{\varepsilon^\prime_\gamma,\rm syn} + j^\prime_{ \varepsilon^\prime_\gamma,\rm ic}$ is the total emissivity.
We neglect electron-positron annihilation, SSA heating of electrons, and diffusive heating and cooling by continuous Thomson scattering \citep{Vurm2009_continuous}.

To solve Equations \eqref{eq:transport}, we employ the Astrophysical Multimessenger Emission Simulator (AMES) \citep{Zhang:2023ewt}.
AMES includes all the processes in Equation \eqref{eq:transport}, as well as attenuation by the extra-galactic background light (EBL).
It provides $n^\prime_{\varepsilon^\prime_\gamma}$ (in units of $\rm erg^{-1}~cm^{-3}$) from $t^\prime = 0$ to  $t^\prime = t^\prime_{\rm dyn}$.
The observed time-averaged flux (in units of $\rm erg~cm^{-2}~s^{-1}$) is then estimated as \citep{Dermer09book} 
\begin{equation}
\begin{split}
\varepsilon_\gamma F_{\varepsilon_\gamma} &= \frac{r_{\rm diss}^2  \Delta r^\prime\Gamma^2}{d_L^2 }\left.\left\langle\varepsilon_\gamma^{\prime2} \frac{n^\prime_{\varepsilon^\prime_\gamma}}{t^\prime_{\rm esc}} \right\rangle \right|_{\varepsilon_\gamma^\prime=\varepsilon_\gamma(1+z)/\Gamma} \rm exp[-\tau_{\rm EBL}(\varepsilon_\gamma)],\\
&=\frac{r_{\rm diss}^2  c\Gamma^2}{d_L^2 }\left.\left\langle\varepsilon_\gamma^{\prime2} n^\prime_{\varepsilon^\prime_\gamma} \right\rangle \right|_{\varepsilon_\gamma^\prime=\varepsilon_\gamma(1+z)/\Gamma} \rm exp[-\tau_{\rm EBL}(\varepsilon_\gamma)],\\
\end{split}
\end{equation}
where 
\begin{equation}
    \left\langle x \right\rangle =\int_0^{t^\prime_{\rm dyn}}\frac{dt^\prime}{t^\prime_{\rm dyn}} x
\end{equation}
is the average of a value $x$ in the calculation time, and $\tau_{\rm EBL}$ is the optical depth of EBL \citep{Gilmore2012EBL}. 

We set following parameters related to the dissipation.
As internal dissipation models, the shock between the jet material and the reconnection-driven turbulence are frequently discussed \citep{Rees1994ApJ...430L..93R,Zhang2011ApJ...726...90Z}, but here we use the fomulation that can be aplied to both.
The nonthermal electron energy densities can be written as
$U^\prime_e = L_{\rm e,iso}/(4\pi r_{\rm diss}^2 \Gamma^2 c)$, and we set the energy injection rate of the electrons as
$\dot U^\prime_{\rm e,inj} = U^\prime_e/t^\prime_{\rm dyn}$. 
In addition, magnetic field energy density is set as $U^\prime_B = \xi_B U_e^\prime$ with a parameter $\xi_B$.

We define $p^\prime_{e,\rm min}$ and $p^\prime_{e,\rm max}$ as the minimum and maximum momentum of the injected electrons.
Based on these definitions, we assume that the differential injection rate of electrons with respect to the momentum (in units of $\rm cm^{-3}~s^{-1}~(eV/c)^{-1}$) is given by
\begin{equation}
\label{eq:dndep}
\begin{split}
\dot n^\prime_{p^\prime_e,\rm diss} 
 &= C_{\rm norm} p^{\prime-p}_e\exp\left(-\frac{p^\prime_e}{p^\prime_{ e,\rm max}}\right) \Theta(p^\prime_e-p^\prime_{e,\rm min}),
\end{split}
\end{equation}
where $p$ is the power-law index, $\Theta$ is the Heaviside step function, and $C_{\rm norm}$ is a normalization factor.
The factor $C_{\rm norm}$ is chosen such that $\int\dot n^\prime_{p^\prime_e,\rm diss} \varepsilon^{\prime}_{e} dp^{\prime}_{e}  = \dot U_{\rm e,inj}^\prime$, where $\varepsilon^{\prime}_{e} = \sqrt{p^{\prime2}_ec^2+m_e^2c^4}$ is the electron energy.
Here, $p^\prime_{e\rm,min}$ is set as a parameter, and $p^\prime_{e\rm,max}$ is defined such that $t^{\prime-1}_{\rm cool}(p^\prime_{e\rm,max})= t^{\prime-1}_{\rm acc}(p^\prime_{e\rm,max})$, where $t^\prime_{\rm acc}$ is the acceleration timescale of electrons.
We assume that the electrons are efficiently accelerated within one cycle of the gyro motion in the dissipation process.
The acceleration timescale as a function of $p_e^\prime$ is given by $t^\prime_{\rm acc} (p_e^\prime) = p_e^\prime/(eB^\prime)$, where the magnetic field is $B^\prime = \sqrt{8\pi U^\prime_B}$.

\section{Result}
\label{sec:result}

Here, we calculate the spectra with AMES for the representative points of ($\Gamma$, $r_{\rm diss}$), labeled (A) to (E) and (a) to (c) listed in Table \ref{tab:param_a2f} and the parameters listed in Table \ref{tab:param}.
The spectra for (a) to (c) are similar to that of (A) to (C).
They are chosen based on the results shown later. 
The classical picture assumes that $\Gamma$ and $r_{\rm diss}$ are connected through the relation $\delta t = r_{\rm diss}/(2\Gamma^2 c)$, where $\delta t$ is the variability timescale in the light curve.
Although $\delta t \sim 100$ s is often adopted for XFs, in this work we explore a broad range of $r_{\rm diss}/(2\Gamma^2 c)$ to account for the observational and theoretical uncertainties in both $\delta t$ and the above relation\footnote{Theoretically, the relation $\delta t = r_{\rm diss}/(2\Gamma^2 c)$ implicitly assumes a shock-dissipation scenario \citep{Bing2019text}, whereas reconnection-driven dissipation does not necessarily follow this relation \citep{Zhang2011ApJ...726...90Z}.
Moreover, although $\delta t$ is often interpreted as the duration of XFs and EEs ($\sim 100$ s) based on the curvature relation discussed in \cite{Liang2006ApJ...646..351L}, it could represent the duration of central engine activity.
The time $\delta t$ that corresponds to $r_{\rm diss}/(2\Gamma^2 c)$ should be the minimum variability timescale.
The light curves of XFs \citep{Peng2014XF_spectra,Zhou2025_170519} and EEs \citep{Sakamoto2011ApJS..195....2S,Kaneko2015MNRAS.452..824K} suggest that shorter variability may be present.
In addition, observationally, the typical XRT count rate for XFs is $\sim 100$ photons s$^{-1}$, making variability on timescales shorter than $0.1$ s uncertain with current observations.}.

The parameters $\xi_B$, $p$, and $p_{\rm e,min}$ listed in Table \ref{tab:param} are uncertain from both theoretical and observational perspectives.
To clarify the dependence on these parameters, Appendixes \ref{app:magjet}, \ref{app:p25}, and \ref{app:pmin20} discuss the magnetized jet case corresponding to $\xi_B = 1$, the soft electron injection case with $p = 2.5$, and the low-$p_{\rm e,min}$ case, respectively. Overall, our qualitative picture remains valid, even if these parameters change.

The identical parameter set is used for both XFs and EEs in this work.
Both XFs and EEs are X-ray emissions with similar luminosities and light-curve shape produced by internal dissipation, and they may therefore share similar parameter values.
The Appendixes also cover the possibility of different parameter sets.

\begin{table}
  \caption{Surveyed parameters}
  \centering
  \begin{tabular}{lcc}
    \hline 
      &$r_{\rm diss}$ & $\Gamma$  \\
       & ($\rm cm$)&   \\
    \hline 
    (A)& $3\times10^{15}$& $30$  \\
       (B)& $3\times10^{15}$& $200$  \\
        (C)& $3\times10^{15}$& $500$  \\
      (a)& $10^{14}$& $80$  \\
       (b)& $10^{14}$& $200$  \\
        (c)& $10^{14}$& $500$  \\
         (D)& $10^{13}$& $150$  \\
          (E)& $10^{13}$& $500$  \\
    \hline
  \end{tabular}
\label{tab:param_a2f}
\end{table}

\begin{table}
  \caption{Fixed parameters }
  \centering
  \begin{tabular}{lccccccccccc}
    \hline 
     &$L_{\rm e,iso}$ & $\xi_B$  & $p^\prime_{e,\rm min}$  & $p$& $d_L$ & $z$ \\
      & ($\rm erg/s$)&    & ($m_ec$)  &   &(Gpc) & \\
    \hline 
      & $10^{50}$& $10^{-1}$  & 200 &2 & 5.0 & 0.8   \\
    \hline
  \end{tabular}
\label{tab:param}
\end{table}

\begin{figure*}\hspace{-0.5cm}
    \begin{tabular}{cc}
      \begin{minipage}[t]{0.5\hsize}
        \centering
        \includegraphics[keepaspectratio, scale=0.48]{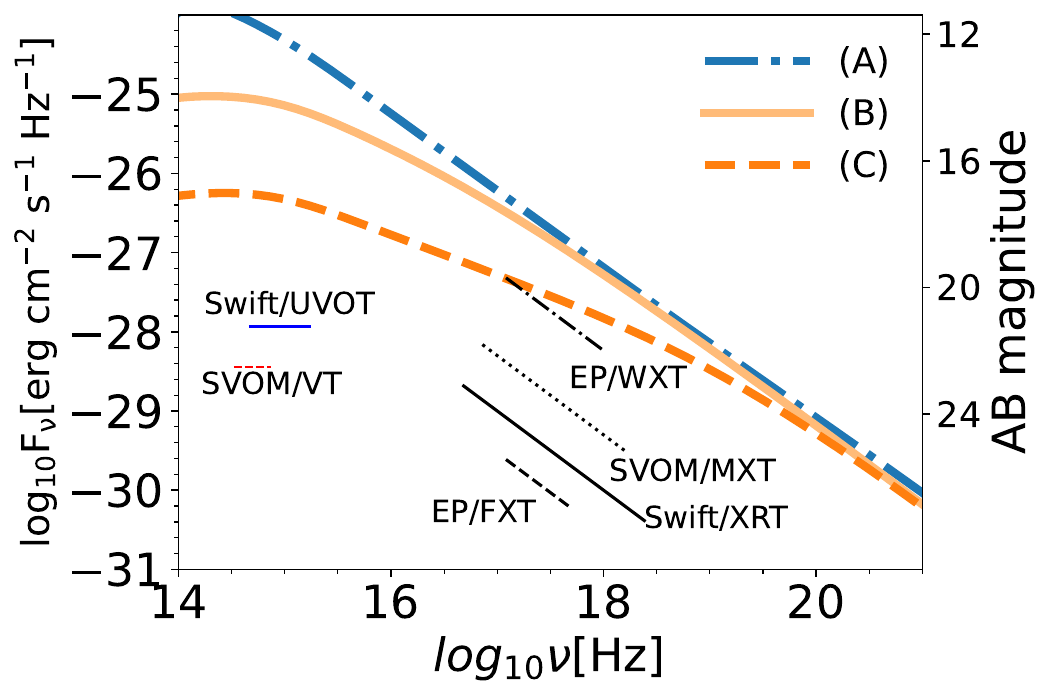}
      \end{minipage}
      
      \begin{minipage}[t]{0.5\hsize}
        \centering
        \includegraphics[keepaspectratio, scale=0.48]{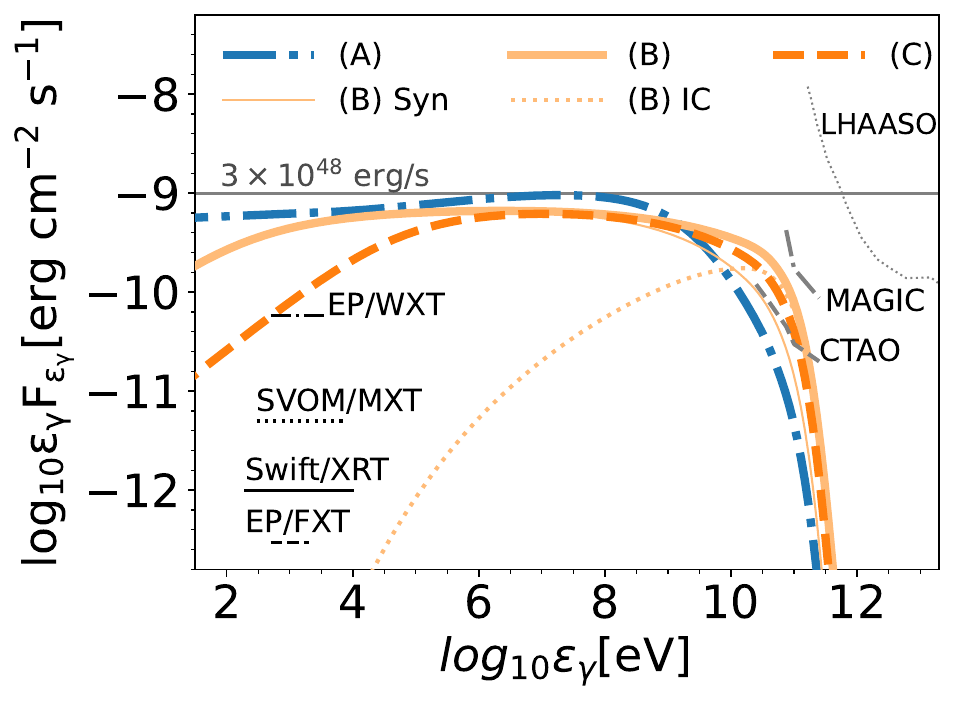}
      \end{minipage}
    \end{tabular}

    \begin{tabular}{cc}\hspace{-0.5cm}
      \begin{minipage}[t]{0.5\hsize}
        \centering
        \includegraphics[keepaspectratio, scale=0.48]{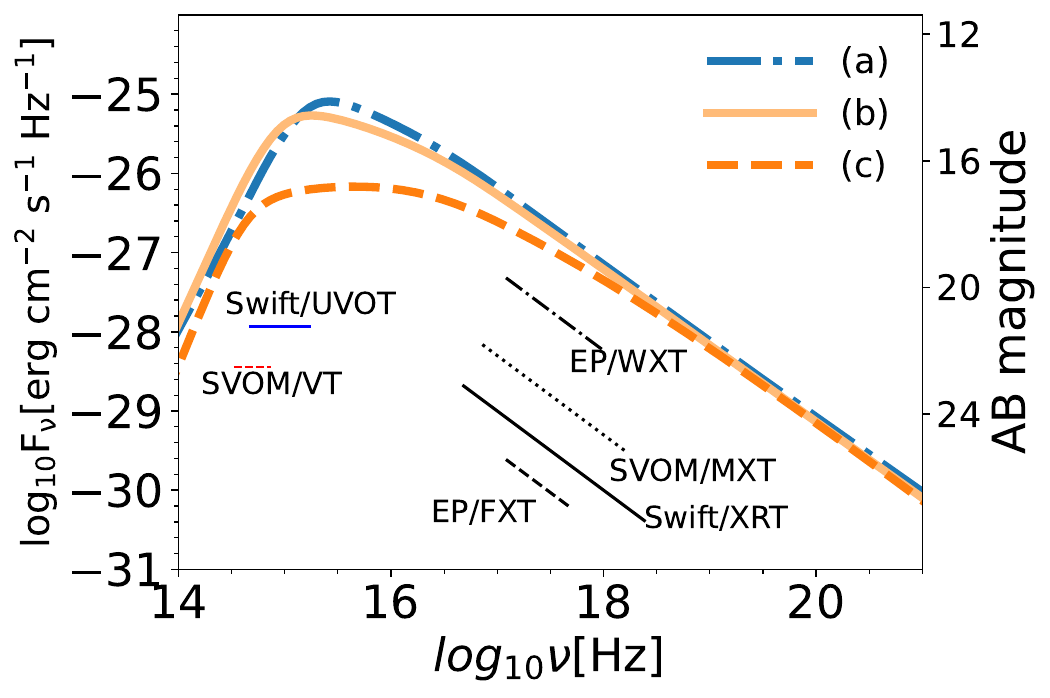}
      \end{minipage}
      
      \begin{minipage}[t]{0.5\hsize}
        \centering
        \includegraphics[keepaspectratio, scale=0.48]{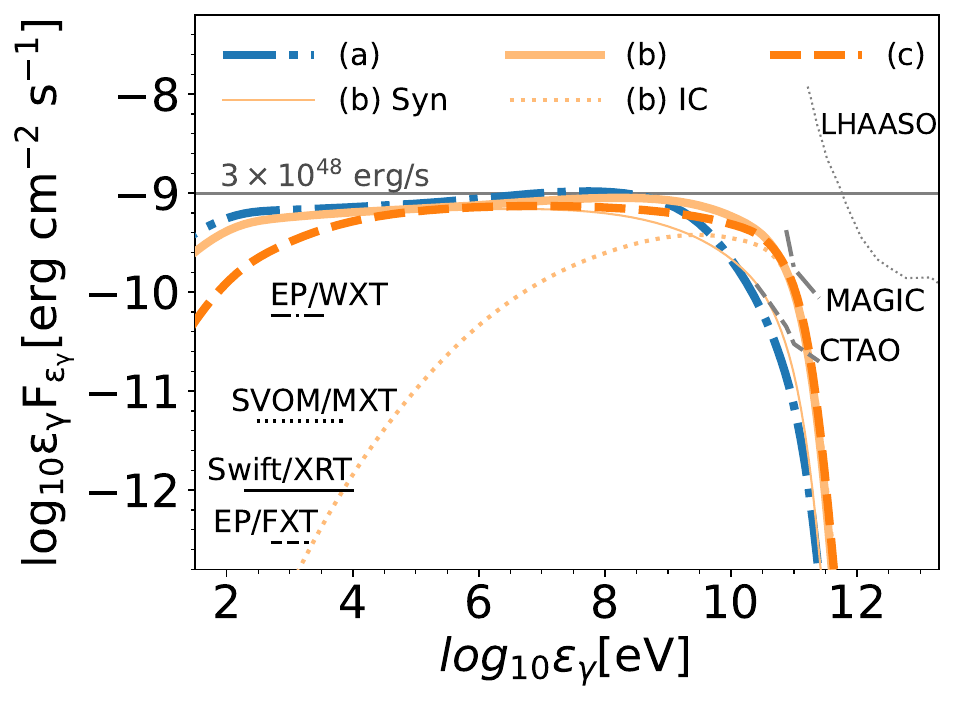}
      \end{minipage}
    \end{tabular}
    
    \begin{tabular}{cc}\hspace{-0.5cm}
      \begin{minipage}[t]{0.5\hsize}
        \centering
        \includegraphics[keepaspectratio, scale=0.48]{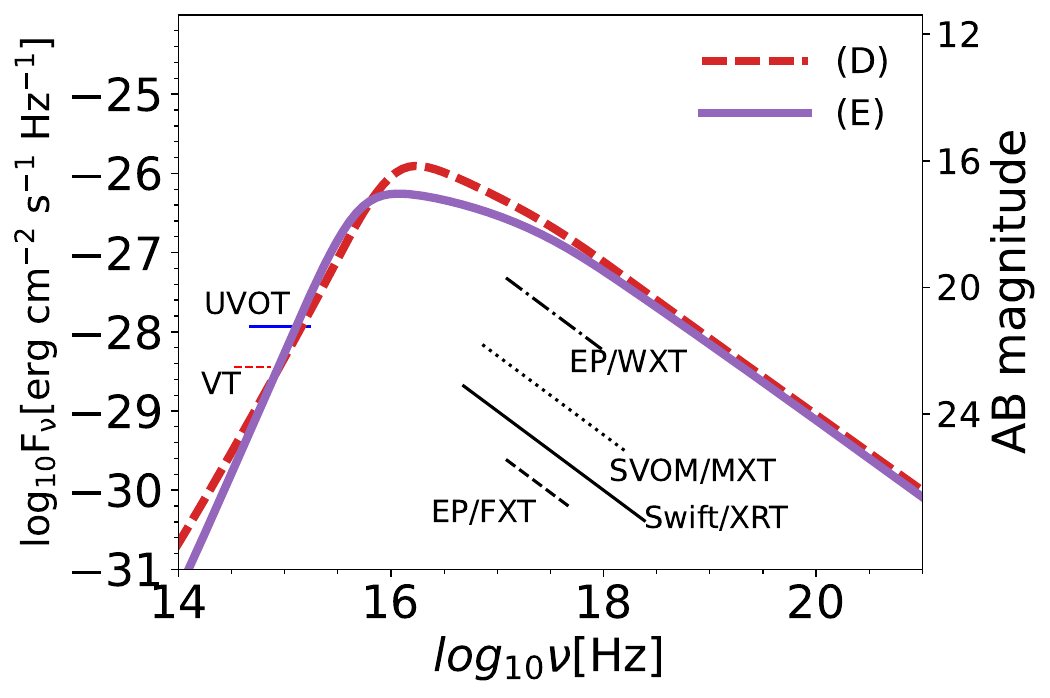}
      \end{minipage}
      
      \begin{minipage}[t]{0.5\hsize}
        \centering
        \includegraphics[keepaspectratio, scale=0.48]{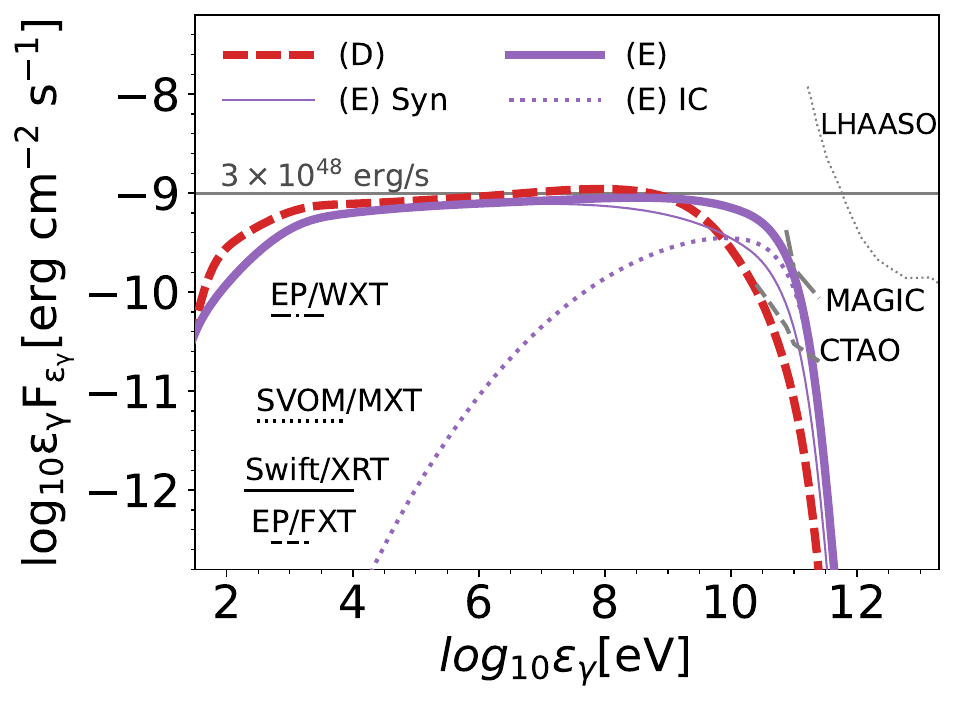}
      \end{minipage}
      
    \end{tabular}
    
    \caption{Spectra for cases (A)-(E) and (a)-(c). Left panels is for $F_\nu$, and right panels is for $\nu F_\nu = \varepsilon_\gamma F_{\varepsilon_\gamma}$.
    The top panels show cases (A)-(C).
    The dotted-dashed blue line, solid light-orange line, and dashed orange line represent cases (A), (B), and (C), respectively. 
    The middle panels show cases (a)-(c).
    The dotted-dashed blue line, solid light-orange line, and dashed orange line represent cases (a), (b), and (c), respectively. 
    The bottom panels show cases (D) and (E).
   In the right panels for cases (B), (b), and (E), the synchrotron and inverse-Compton components are plotted as thin solid and dotted lines, respectively, in the corresponding colors.
    The dashed red line, solid purple line correspond to cases (D) and (E), respectively.
    The horizontal gray lines in right panels ($\varepsilon_\gamma F_{\varepsilon_\gamma}$ plots) indicate the typical luminosity of XFs and EEs, $3\times10^{48}$ erg/s.
    The other thin lines indicate the detection limits for 300-second transients: the dashed red lines are for SVOM/VT \citep{Atteia2022_SVOM}, the solid blue lines are for Swift/UVOT \citep[][1000 s white sensitivity $\times \sqrt3$]{Gehrels2004_swift}, the dotted black lines are for SVOM/MXT \citep{Atteia2022_SVOM,Gotz2014MXTsensitivity}, the solid black lines are for Swift/XRT \citep{Gehrels2004_swift,Burrows2005XRTsensitivity}, the dotted-dashed black lines are for WXT on Einstein Probe \citep[log-log interpolation between 100 s and 1000 s sensitivities]{Yuan2022EP}, the dashed black lines are for FXT on Einstein Probe \citep[][1500 s sensitivity in 0.5-2 keV $\times 5$]{Zhang2022EP}, the dashed gray lines are for CTAO \citep{Hofmann2023_CTAO}, the dotted-dashed gray lines are for MAGIC \citep{Aleksic2016MAGIC,Fioretti2019MAGICsensitivity}, and the dotted-dashed gray lines are for LHAASO \citep[][1 year sensitivity $\times\sqrt {1~\rm year/300 ~s}$]{diSciascio2016LHAASO}. The CTAO and MAGIC sensitivity lines end at 250 GeV in our plots following Figure 22 in \cite{Hofmann2023_CTAO} and \cite{Fioretti2019MAGICsensitivity}, but actually it extends to 100 TeV. }
    \label{fig:spectra}
  \end{figure*}

\subsection{Parameter Survey}
\label{param_survey}

Figure \ref{fig:spectra} shows the resultant spectra.
The left and right panels are for $F_\nu = F_{\varepsilon_\gamma} h$ in units of $\rm erg~cm^{-2}~s^{-1}~Hz^{-1}$ and for $\varepsilon_\gamma F_{\varepsilon_\gamma}$ in units of $\rm erg~cm^{-2}~s^{-1}$, respectively.
The frequency is defined as $\nu= \varepsilon_\gamma/h$.
They are all adjusted to have the typical X-ray luminosity of XFs and EEs: $0.3-1\times 10^{49}$ erg/s.
For cases (A)-(C) and (a)-(c), the UV ($\sim 1$ eV) emission is detectable by Swift/UVOT \citep{Gehrels2004_swift} and SVOM/VT \citep{Atteia2022_SVOM}.
The spectra for (D) and (E) show significant breaks at $\lesssim$0.1 keV ($\sim 10^{16}$ Hz) made by SSA, preventing Swift/UVOT and SVOM/VT from detecting\footnote{The left bottom panel of Figure \ref{fig:spectra} shows the spectra for cases (D) and (E) overlapping with the UVOT detection limit.
However, the limit shown corresponds to the sensitivity of the UVOT white band. We confirm that the integrated fluxes over the white band are below the UVOT white-band sensitivity.}.
Moreover, for cases (B), (C), (b), (c), and (E), the very-high-energy (VHE) gamma rays ($\sim 10$ GeV) can be detected by CTAO \citep{Hofmann2023_CTAO}.
The highest energy of photons for these cases are determined by the EBL cut off at around 100 GeV.
In contrast, the spectra for (A), (a) and (D) exhibit breaks below $\sim$1 GeV made by $\gamma\gamma$ absorption, making it difficult to detect in VHE gamma rays.
Other VHE gamma ray detecotors such as MAGIC \citep{Aleksic2016MAGIC}, H.E.S.S.\citep{Aharonian2006HESS}, VERITAS \citep{Weekes2002VERITAS}, LHASSO \citep{Cao2019LHAASO} cannot detect emissions from all cases because they are sensitive at $>100$ GeV. 
As their representative, the sensitivity of MAGIC is plotted in Figure \ref{fig:spectra}.

\begin{figure*}[ht!]
\centering
\includegraphics[keepaspectratio, scale=0.7]{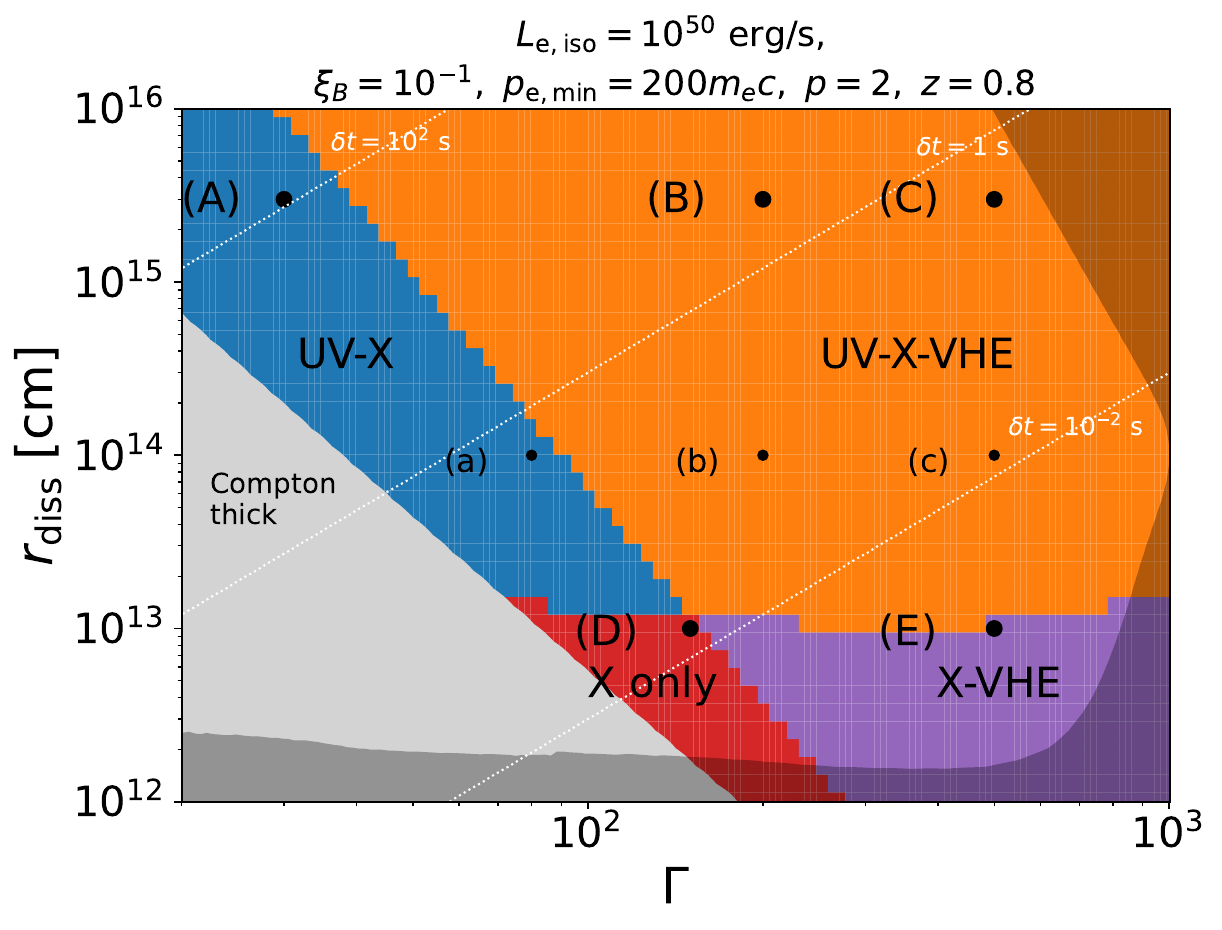}
\caption{Detectabilities on the $r_{\rm diss}$-$\Gamma$ plane.
The blue region indicates that UV emission associates with XFs or EEs but no VHE gamma-ray emission.
The orange region indicates that both UV (Swift/UVOT) and VHE (CTAO) emissions associate with XFs or EEs.
The red region indicates that neither UV nor VHE emission associates with XFs or EEs.
The purple region indicates that VHE emission associates with XFs or EEs but no UV emission.
The black shaded region is where the photon index at the X-ray band is too hard for the XFs and EEs.
The gray region is not considered because the Thomson optical depth exceeds unity and the emission there is subphotospheric.
The thin white dotted line represents where the $\delta t = r_{\rm diss}/(2\Gamma^2 c)$ is constant.
}
\label{fig:r-Gamma}
\end{figure*}

The parameter survey results in $r_{\rm diss}-\Gamma$ plane is shown in Figure \ref{fig:r-Gamma}.
For high $\Gamma$ and large $r_{\rm diss}$ in the orange region, we expect both UV and VHE emissions to be associated with X-rays.
In this region, including cases (B), (C), (b) and (c), SSA and $\gamma\gamma$ absorption do not affect the UV band or the VHE gamma rays.
For high $\Gamma$ and small $r_{\rm diss}$ in the purple region, including cases (E), VHE gamma rays can be as luminous as X-rays, while UV emission is absorbed by SSA.
For intermediate $\Gamma$ and large $r_{\rm diss}$ in the blue region, including case (A) and (a), UV emission can avoid SSA and may be associated with the X-rays, whereas VHE gamma rays are still absorbed by $\gamma\gamma$.
For intermediate $\Gamma$ and small $r_{\rm diss}$ in the red region, including case (D), we expect only X-rays, because the UV emission and VHE gamma rays are fully absorbed by SSA and $\gamma\gamma$ processes, respectively.

The black shaded region indicates areas where the spectral index is harder than 1.5.
This is inconsistent with that of typical XFs or EEs.
In addition, for low $\Gamma$ and small $r_{\rm diss}$ in the gray region, the Thomson optical depth, $\tau_T$, exceeds unity \citep{Beniamini2016-photospheric_flare}.
We do not consider the parameter space in this paper.
This criterion is equivalent with the limit on the Lorentz factor of EEs given by \cite{Matsumoto2020MNRAS.493..783M}.
The photosperic emission in this regime is also discussed in Section \ref{sec:photospheric}.

The MeV gamma-ray luminosity can be comparable to the X-ray luminosity.
For a typical luminosity of $\sim 10^{49}$ erg/s and a typical redshift of $z \sim 1$-$3$, such emission cannot be detected by Fermi/GBM \citep{Axelsson-Fermi} or Swift/BAT \citep{Gehrels2004_swift} due to their sensitivity limits.
However, EEs in MeV gamma rays are detected for some short GRBs \citep{Sakamoto2011ApJS..195....2S,Kaneko2015MNRAS.452..824K}.
Our fiducial calculation for nearby events such as $z<0.25$ can explain the observed flux of $\sim10^{-8}\rm ~erg~cm^{-2}~s^{-1}$ at MeV range. 
If they have comparable GeV flux as cases (B), (C), (b), (c), and (E), Fermi/LAT can detect the event.
The Fermi/LAT detectability of such events is discussed in Section \ref{sec:discussion}.

\subsection{Case study: GRB 060926}
\label{sec:case_UV-X}
UV flares clearly associated with XFs have been reported for a few GRBs, although comprehensive cross-correlation analyses have not yet been carried out.
The light curves presented in \cite{Yi2017_UVF} for GRB 060926, \cite{Becerra2021_GRB1803325} for GRB 180325A, and \cite{Jin2023_GRB220101} for GRB 220101A show a clear correlation between the X-ray and UV bands.

\begin{figure}\hspace{-1cm}
        \centering
        \includegraphics[keepaspectratio, scale=0.52]{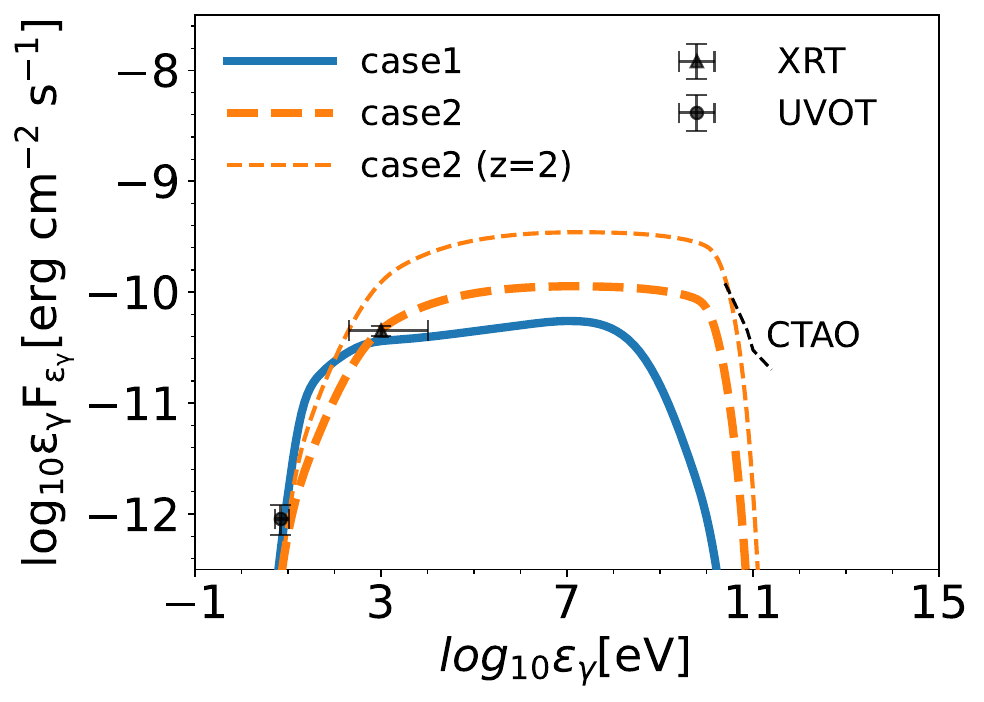}
    \caption{Spectrum of the flare in GRB 060926 at $T - T_0 \sim 10^{2.5}$ s. The thick solid blue line and thick dashed orange line show the spectra for cases 1 and 2 in Table \ref{tab:param_case}, respectively. The thin dashed line shows the spectrum for case 2 at $z = 2$. The triangular and circular points represent the observed data from Swift/XRT and UVOT, respectively. The thin dashed black line indicates the detection limit of CTAO same as in Figure \ref{fig:spectra} \citep{Hofmann2023_CTAO}.}
    \label{fig:spectra_case_UV-X}
\end{figure}

We focus on GRB 060926 here because its flare is comparable in luminosity to a typical X-ray flare, whereas the others are as luminous as the prompt emissions.
The parameters chosen for explaining the observation are shown in Table \ref{tab:param_case}, and calculated spectra are shown in Figure \ref{fig:spectra_case_UV-X}.
The parameters not listed in Table \ref{tab:param_case} follows Table \ref{tab:param}.

Table \ref{tab:param_case} shows two cases: case 1 and case 2.
Case 1 corresponds to a low $\Gamma=100$, while case 2 corresponds to a high $\Gamma=1000$.
To reproduce the observed UV photons avoiding SSA, both require a relatively large dissipation radius of $r_{\rm diss} \gtrsim 10^{13}$ cm.
At low $\Gamma$, the $\gamma\gamma$ absorption has important role, while it is inefficient at high $\Gamma$ as shown in Figure \ref{fig:r-Gamma}.
Thus, VHE gamma-rays $\gtrsim10$ GeV appear only for case 2. 
Although it cannot be detected by CTAO due to the suppression by EBL attenuation, the event with the parameter case 2 at $z\lesssim2 $ could be detected by CTAO as shown by the thin dashed line in Figure \ref{fig:spectra_case_UV-X}. 
From this event, we can clearly see that simultaneous observations of the UV and X-ray flares with CTAO are important to constrain the value of $\Gamma$.

\begin{table}
  \caption{Parameters for GRB 060926 }
  \centering
  \begin{tabular}{lcccccccccc}
    \hline 
      &$L_{\rm e,iso}$ & $r_{\rm diss}$ & $\Gamma$ & $d_L$ & $z$ \\
      & ($\rm erg/s$)   & (cm) & & (Gpc)&\\
      \hline 
    case 1 & $10^{50}$& $1.6\times10^{13}$ & $10^{2}$&  27 & 3.2   \\
     case 2 & $3\times 10^{50}$& $1.6\times10^{13}$ & $10^{3}$&   27 & 3.2  \\
    \hline
  \end{tabular}
\label{tab:param_case}
\end{table}

\section{Constraints and detection prospects}
This section expect detections associated with XFs and EEs for the future.
\label{sec:detection}
The UV association for GRB 060926, as discussed in Section \ref{sec:case_UV-X}, indicates that $r_{\rm diss}$ exceeds $10^{13}$ cm for the GRB.
However, to analyze the majority of GRBs with flares, we need to examine the cross-correlation between UV and X-ray flares.
Since UV flares are reported in 10-30\% of long GRBs \citep{Swenson2013UVF,Swenson2014-XF,Yi2016ApJS..224...20Y,Yi2017_UVF}, the cross-correlation analysis is important for understanding the XFs and EEs.

The survey of VHE gamma-ray flares is important to constrain $\Gamma$.
To observe a GRB with CTAO, an alert localization of $\sim \rm 1~degree$ is required \citep{Hofmann2023_CTAO}.
This means that the GRB must be triggered by Swift/BAT \citep{Gehrels2004_swift} or SVOM/ECLAIR \citep{Atteia2022_SVOM} in the current situation.
The detection rate of GRBs by Swift/BAT is roughly $80\ \rm yr^{-1}$ \citep{Lien2016ApJ...829....7L}, and SVOM/ECLAIR is expected to achieve a similar frequency \citep{Atteia2022_SVOM}.
Additionally, as shown in the right panel of Figure \ref{fig:spectra}, $z \sim 0.8$ represents the detection horizon of VHE gamma rays for CTAO for the flare with typical X-ray luminosity $\sim 10^{49}$ erg/s.
Among the events with measured redshifts, about $\sim 10\%$ are from $z < 0.8$ \citep{Lien2016ApJ...829....7L,Hao2020redshift_dist}.
From the above arguments, we expect about $\sim 8~\rm yr^{-1}$ of $z < 0.8$ GRBs to be detected by Swift/BAT or SVOM/ECLAIR.
Considering that 30\% of GRBs have XFs or EEs, BAT and ECLAIR can detect about $\sim 3~\rm yr^{-1}$ of $z < 0.8$ GRBs with flares.
Assuming that roughly 10\% of GRBs can be observed by CTAO with a good condition \citep{Inoue2013CTA_dutycycle}, a VHE flare can be detected about once every $\sim 3$ years with current detector systems.
Such a detection with CTAO would reveal the Lorentz factor of the flaring jet.

Our calculation can also be applied to EEs.
However, the smaller number of EEs compared with XFs reduces the expected detection rate.
EEs appear in about 50\% of short GRBs (sGRBs) \citep{Kisaka2017ApJ...846..142K}, and sGRBs are detected by Swift/BAT at a rate of $\sim8\ \rm yr^{-1}$ \citep{Lien2016ApJ...829....7L}.
Therefore, the rate of Swift/BAT sGRBs with EEs is $\sim4\ \rm yr^{-1}$, and the expected detection rate with CTAO becomes $\sim0.04\ \rm yr^{-1}$.
Thus, detecting such events in VHE gamma rays would require $\sim 20$ yr.

VHE gamma-ray detectors currently conducting GRB follow-up campaigns, such as MAGIC \citep{Aleksic2016MAGIC}, H.E.S.S. \citep{Aharonian2006HESS}, VERITAS \citep{Weekes2002VERITAS}, and CTA LST-1 \citep{Abe2023CTALST1}, can potentially detect emission associated with XFs and EEs.
To constrain $\Gamma \sim 100$, GRBs with redshift $z \lesssim 0.5$ are required to avoid significant attenuation by the EBL.
The event rate for $z \lesssim 0.5$ is $\sim 3$-$5$\% of all GRBs \citep{Hao2020redshift_dist}.
Considering that XFs or EEs are associated with $\sim 30$\% of GRBs, we expect $\sim 1$-$2$ XFs or EEs from $z < 0.5$ per 100 GRBs without redshift measurements.
Since these facilities have collectively observed more than 100 GRBs with delays shorter than 1000 s after the prompt emission \citep{Abe2025MAGICGRB,Ribeiro2023VERITASGRB,Cornejo2025HESSGRB}, such observations might already provide meaningful constraints.
A careful analysis of the upper limits, particularly accounting for the timing of XFs and EEs, is required.

For nearby events, Fermi/GBM and Fermi/LAT can detect MeV and GeV gamma rays from XFs and EEs.
If MeV gamma rays are detected with XFs, it is difficult to distinguish between hundred-second-long prompt emission and XFs based on the light curve alone.
On the other hand, EEs may be distinguished from prompt emission in the light curve.
For this reason, here we focus on the detection rate of EEs in MeV and GeV gamma rays.
According to \cite{Kaneko2015MNRAS.452..824K}, Fermi/GBM and Swift/BAT detected 28 EE events with fluxes of $\gtrsim 10^{-8}\ \rm erg\ cm^{-2}\ s^{-1}$ over $\sim 8$ years.
This flux is close to the minimum value observed by LAT in the 100-1000 s interval after the prompt emission of short GRBs \citep{Ajello2019ApJ...878...52A}.
This implies that EEs detectable by Fermi/LAT at $\sim 1$ GeV occur at a rate of $\sim 3.5~\rm yr^{-1}$ for the orange and purple regions in Figure \ref{fig:r-Gamma}.
Since Fermi/LAT covers about one-sixth of the sky \citep{Ajello2019ApJ...878...52A}, it can detect EEs at a rate of $\sim 0.6~\rm yr^{-1}$, which is similar to the expected detection rate of flares by CTAO.

Actually, GeV gamma rays have been detected by Fermi/LAT from several short GRBs over $\sim 10$ years of observations \citep{Ajello2019ApJ...878...52A}, although it remains unclear whether these emissions are associated with EEs. 
We do not model them here because redshifts are not determined for most of the events, and the one event with a known redshift shows a light curve that is clearly different from typical EEs.
Above arguments demonstrate that Fermi/LAT observations and event by event broadband modelings are also important for constraining the $r_{\rm diss}$-$\Gamma$ plane.

In the near future, nanosatellite constellations such as CAMEROT \citep{Werner2018CAMELOT} and HERMES \citep{Fuschino2019HERMES} will improve the detection frequency of events with localization accuracy of about $\rm 1~degree$.
This will increase the chance for CTAO to detect emissions associated with XFs and EEs by a factor of $\sim$3.
Therefore, VHE flares could be observed at a rate of $\sim 1~\rm yr^{-1}$ with these detectors.

\section{discussion} 
\label{sec:discussion}
In this section, we discuss expected contamination, application, and extensions of our work.  

\subsection{Possible contamination: External inverse Compton at the forward shock}
\label{sec:lightcurve}
The VHE gamma-ray emission expected in this paper may be contaminated by an emission component from the forward shock (FS), which produces the afterglow.
XFs and EEs inevitably provide X-rays that serve as seed photons for external inverse Compton (EIC) scattering at the FS, and the resulting EIC emission in GeV-TeV energies appears at a similar epoch \citep{Wang2006_GeVTeV_XF,Murase2011XFEE_EIC,Zhang2021_EEEIC}.
Moreover, its flux can reach comparable to those of XFs and EEs at a maximum \citep{Wang2006_GeVTeV_XF}, leading to the possible significant contamination. 

Below we provide a quantitative estimate of the EIC luminosity, $L_{\rm EIC}$.  
The luminosity can be written as $L_{\rm EIC} \sim \eta_{\rm EIC} L_{e,\rm iso,FS}$, where $\eta_{\rm EIC}$ is the EIC efficiency and $L_{e,\rm iso,FS} \sim \epsilon_e E_{\rm iso} t^{-1} \sim 10^{48} \epsilon_{e,-2} E_{\rm iso,53} t_3^{-1}~{\rm erg~s^{-1}}$ \citep{Wang2006_GeVTeV_XF} is the electron luminosity at the FS.  
Here, $\epsilon_e$ is the fraction of energy carried by electrons, $E_{\rm iso}$ is the isotropic-equivalent energy of the jet, and $t$ is the time after the prompt emission.  
Typically, $L_{e,\rm iso,FS}$ is comparable to the typical luminosity of XFs and EEs, $L_{e,\rm iso}$.  
The efficiency, $\eta_{\rm EIC}$, for electrons with Lorentz factor $\gamma_{e,\rm FS}^\prime$ can be roughly estimated as  
$\eta_{\rm EIC} \sim f_{\rm ene}(\gamma_{e,\rm FS}^\prime)\, t_{\rm dyn}^\prime / t_{\rm EIC}^\prime(\gamma_{e,\rm FS}^\prime)$,  
where $t_{\rm EIC}^\prime(\gamma_{e,\rm FS}^\prime)$ is the EIC cooling timescale and $f_{\rm ene}(\gamma_{e,\rm FS}^\prime)$ is the fraction of electron energy carried by electrons at $\gamma_{e,\rm FS}^\prime$.

The electron distribution prior to EIC is determined by synchrotron cooling.  
Typical afterglow at $t = 1000$ s is in the slow-cooling regime, shown by \citep{Murase2011XFEE_EIC}
\begin{equation}
\label{eq:gamma_m_gamma_c}
\begin{split}
    &\gamma^\prime_{m}\sim 2.3\times10^2\epsilon_{e,-2} f_{e}^{-1}(g_{2.4}/g_{p_{\rm FS}})E_{\rm iso,53}^{1/8}n_0^{-1/8} t_3^{-3/8}(1+z)^{3/8}\\
    &< \gamma^\prime_{c,\rm syn}\sim 6.2\times10^4\epsilon_{B,-3}^{-1}E_{\rm iso,53}^{-3/8}n_0^{-5/8}t_3^{1/8} (1+z)^{-1/8},
\end{split}
\end{equation}
where $\gamma^\prime_{m}$ is the minimum Lorentz factor of electrons, $\gamma^\prime_{c,\rm syn}$ is an electron Lorentz factor that satisfies $t_{\rm syn}^\prime(\gamma^\prime_{c,\rm syn})=t_{\rm dyn}^\prime$, $n_0$ is the ISM density, $f_e$ is the fraction of nonthermal electrons, $g_{p_{\rm FS}}= (p_{\rm FS}-1)/(p_{\rm FS}-2)$, $p_{\rm FS}$ is the power-law index at the FS, and  All these parameters are defined at the FS.
Here, we consider only the slow-cooling regime.

In this regime, we obtain $f_{\rm ene} (\gamma_{e,\rm FS}^\prime) \sim (\gamma_{e,\rm FS}^\prime/\gamma^\prime_{m})^{2-p_{\rm FS}}$ for $\gamma^\prime_{m}< \gamma^\prime_{e,\rm FS}< \gamma^\prime_{c,\rm syn}$.
This relation, together with $t_{\rm EIC}^\prime(\gamma_{e,\rm FS}^\prime)\propto \gamma^{\prime-1}_{e,\rm FS}$ suggests that $\eta_{\rm EIC}$ peaks at $\gamma_{e,\rm FS}^\prime = \gamma^\prime_{c,\rm syn}$ for $2<p_{\rm FS}<3$.
After EIC, the peak shifts to $\gamma^\prime_{e,\rm FS} = \gamma^\prime_{c} = \gamma^\prime_{c,\rm syn} (1+Y)^{-1}$, where $Y = U_{\rm seed}^\prime/U_{B\rm,  FS}^\prime = (\epsilon_e/\epsilon_B)(L_{e,\rm iso}/L_{e,\rm iso,FS})$ is the Compton-Y parameter, $U_{\rm seed}^\prime$ is the energy density of seed photons, and $U_{B,\rm FS}^\prime$ is the magnetic-field energy density at the FS, $\epsilon_B$ is the magnetic energy fraction at the FS.
Then, the electrons with $\gamma_c$ dominate GeV-TeV emission \citep{Murase2011XFEE_EIC}.
Substituting $\gamma_{e,\rm FS}^\prime = \gamma_c^\prime$ and using $t_{\rm EIC}^{\prime -1} (\gamma_c^\prime) =  Yt_{\rm syn}^{\prime -1}(\gamma_{c,\rm syn})/(1+Y)$, we obtain $\eta_{\rm EIC} \sim f_{\rm ene} (\gamma_{\rm c}^\prime)Y/(1+Y)$, and thus $L_{\rm EIC} \sim (\gamma_{\rm c,\rm syn}^\prime/\gamma^\prime_{m})^{2-p_{\rm FS}} (1+Y)^{p_{\rm FS}-3}(\epsilon_e / \epsilon_B) L_{e,\rm iso}$.
From this relation, the EIC emission becomes as luminous as XFs and EEs when $(\gamma_{\rm c,\rm syn}^\prime/\gamma^\prime_{m})^{(2-p_{\rm FS})} (1+Y)^{3-p_{\rm FS}} (\epsilon_e / \epsilon_B)\gtrsim 1$.
This condition is rewritten as
\begin{equation}
\begin{split}
     &\gamma_{\rm c,\rm syn}^\prime/\gamma^\prime_{m} \\ 
    &\lesssim  ( \epsilon_B/ \epsilon_e )^{\frac{1}{(p_{\rm FS} -2)}}[1+(\epsilon_e/\epsilon_B)(L_{e,\rm iso}/L_{e,\rm iso,FS})]^\frac{p_{\rm FS}-3}{p_{\rm FS}-2},
\end{split}
\end{equation}
for $ \gamma_{\rm c,\rm syn}^\prime/\gamma^\prime_{m}>1$ (slow-cooling).

This condition are typically satisfied for $t\sim 1000$ s \citep{Murase2011XFEE_EIC}, leading to the significant EIC emission at the FS. However, even if this is significant, we can decontaminate as follows.
The EIC light curve rises with a time delay of $R_{\rm FS}/(2c\Gamma_{\rm FS}^2) > r_{\rm diss}/(2c\Gamma^2)$ relative to the X-ray flare, where $R_{\rm FS}$ and $\Gamma_{\rm FS}$ are the radius and Lorentz factor of the FS \citep{Beloborodov2014_EIC}.
In addition, the tail of the EIC light curve becomes broader than that of the seed photons due to high-latitude emission \citep{Beloborodov2014_EIC}.
In contrast, the VHE gamma rays from our scenario should closely follow the X-ray light curve.
Although a fully consistent light-curve calculation is beyond the scope of this work, these two components can, in principle, be distinguished by comparison of VHE light curves with X-ray ones.

\subsection{Potential application: Shallow-decay phase and plateau emission}
\label{sec:shallow}
The shallow-decay phase in long GRBs and the plateau emission in short GRBs are X-ray components characterized by relatively flat light curves \citep[e.g.,][]{Nousek2006ApJ...642..389N,Kisaka2017ApJ...846..142K}.
The shallow-decay phase appears at $\sim 10^3$-$10^4$ s after the prompt emission with a luminosity of $\sim 10^{46}$-$10^{48}$ erg/s \citep{Nousek2006ApJ...642..389N,Zhang2006XAG,Dainotti2010relation,Bernardini2012Shallow_lumi}, whereas the plateau emission appears at $\sim 10^4$-$10^5$ s with a luminosity of $\sim 10^{45}$-$10^{47}$ erg/s \citep{Gompertz2013plateau,Kisaka2017ApJ...846..142K}.
Their origins are still under debate, but they could arise from internal dissipation of late jet components \citep{Kumar2008engine_fallback,Ghisellini2009late-prompt,Lyons2010magnetarPlateau}, although the variability properties of their light curves remain unclear.
Therefore, the discussion in this paper may also be applicable to these components.

However, the X-ray luminosities of these components are lower than those of XFs and EEs.
Since the UV and VHE gamma-ray luminosities are expected to be comparable to the X-ray luminosity at maximum, detecting them from $z \gtrsim 1$ is challenging based on our model.
Therefore, we must wait for nearby events to detect broadband emission associated with these components.

\subsection{Future extension of this study}
\label{sec:extension}
In this work, we ignore the contribution from following components shown in this subsection. 
It is possible to extend our work to include them.

\subsubsection{Physics inside the photosphere}
\label{sec:photospheric}
XFs and EEs can originate from the photosphere \citep{Rees2005ApJ...628..847R,Murase2011XFEE_EIC,Peng2014XF_spectra,Beniamini2016-photospheric_flare,Peng2024_EEphotosphere,Zhou2025_170519}, gray region in Figure \ref{fig:r-Gamma}.
Inside the photosphere, nonthermal dissipation is more complicated because it is mediated by radiation
\citep{Thompson2007_thermalization,Zrake2019_subphoto_turbulence,Lundman2019pair_subshock}.
Further study is required to model XFs and EEs arising from subphotospheric dissipation while treating electromagnetic processes and hydrodynamics self-consistently
\citep{Giannios2006_photosphere,Vurm2009_continuous,Beloborodov2017_subshock,Lundman2019pair_subshock,Zrake2019_subphoto_turbulence}.
Moreover, SSA, Compton heating of electrons, and electron-positron annihilation may affect the photon spectra in the gray region
\citep{Vurm2009_continuous,Beloborodov2010_collisional}.
Calculations including all these processes are left for future work.

\subsubsection{Cocoon photons}
\label{sec:cocoon}
The interaction between the prompt jet and the progenitor star forms the structure called ``cocoon" \citep{Mezaros2001ApJ...556L..37M,Matzner2003MNRAS.345..575M,Bromberg2011ApJ...740..100B,Nakar2017ApJ...834...28N,Gottlieb2020MHD,Gottlieb2021hydro,Hamidani2021Jetprop}.
The cocoon expands after the jet breakout \citep{Hamidani2023escapeCoc}.
The expanded cocoon component can affect the emission of multi-wavelength flares or EEs.

If $r_{\rm diss}$ lies within the cocoon radius, external photons from the cocoon can enter the dissipation region and serve as seed photons for inverse Compton emission and $p\gamma$ process \citep{Toma2009ApJ...707.1404T,Kimura2019ApJ...887L..16K,Mei2022Natur.612..236M,Matsui2023ApJ...950..190M,Matsui2024}.

The cocoon effect is typically significant for $r_{\rm diss} < 10^{13}$ cm and $\Gamma > 200$, roughly overlapping with the purple region in Figure \ref{fig:r-Gamma} \citep{Matsui2023ApJ...950..190M,Matsui2024}.
However, the cocoon energy strongly depends on the luminosity and magnetic field of the prompt jet \citep{Nakar2017ApJ...834...28N,Gottlieb2020MHD,Gottlieb2021hydro,Hamidani2023escapeCoc}, and the strength of the effect depends on the time after the main burst \citep{Kimura2019ApJ...887L..16K,Matsui2023ApJ...950..190M,Matsui2024}, which reduces the simplicity of the picture.
This effect needs to be considered on an event-by-event basis.
A detailed calculation including the cocoon effect is beyond the scope of this work.

\subsubsection{Hadronic emission associated with X-ray flares and extended emission}
\label{sec:hadron}

The dissipation could accelerate protons or heavier nuclei in the jet \citep{Waxman1995GRBCR,Vietri1995GRBCR}.
These dissipated hadrons can emit gamma rays through inelastic nuclei-photon collisions ($p\gamma$ process for protons) and nuclei-nuclei collision ($pp$ process for protons)\citep{Waxman1997PhRvL..78.2292W,Bottcher1998_GRBCR_pgamma,Asano2007_hadroncascade,Asano2009_hadroncascade2,Murase:2011cx}.
Spectral calculations with dissipated hadrons are beyond the scope of this discussion.
However, our calculations may mimic the hadronic results for the following reasons:
First, the photons from the hadronic process are high energy, greater than or equal to 1 GeV, at the jet rest frame.
These photons are easily absorbed by low-energy photons, creating copious electron-positron pairs. 
These pairs produce synchrotron emissions.
This cascading process converts the jet energy into leptonic energy.
This may lead to similar energetics as a purely leptonic process, although the spectra should be modified \citep{Asano2007_hadroncascade,Asano2009_hadroncascade2,Murase:2011cx}. Alternatively, if the Lorentz factor is very large, cascades become less efficient, in which nuclear synchrotron emission or secondaries from the Bethe-Heitler process can be relevant~\citep{Murase:2010va}. 

High-energy neutrino observations provide another way to constrain the $r_{\rm diss}$-$\Gamma$ plane.
For low $\Gamma < 100$ in the blue and red regions of Figure \ref{fig:r-Gamma}, photons produced within the dissipation region can participate in the $p\gamma$ process, leading to neutrino production in the TeV-PeV energy range \citep{Murase2006flareneutrinos}. 
If XFs and EEs are from photosphere, not only $p\gamma$ interactions but also $pp$ interactions can also be relevant~\citep{Murase:2008sp}. 

For high $\Gamma > 100$ with small $r_{\rm diss} < 10^{13}$ cm in the purple region of Figure \ref{fig:r-Gamma}, external photons from the cocoon may enhance the $p\gamma$ process \citep{Matsui2023ApJ...950..190M,Matsui2024}.
Meanwhile, the cocoon photons may kill VHE photons through $\gamma\gamma$ absorption.
A fully consistent numerical calculation and parameter survey covering these multi-messenger signals is left for future work.

\section{Conclusion} 
\label{sec:conclusion}
We calculate the multi-wavelength flux associated with XFs and EEs over a broad range $r_{\rm diss}$ and $\Gamma$ in Table \ref{tab:param_a2f}, using the parameters listed in Table \ref{tab:param}.
The spectra shown in Figure \ref{fig:spectra} suggest the possible simultaneous detection of UV, X-ray, and VHE gamma-ray emission 100-1000 s after the prompt emission of GRBs, associated with XFs or EEs.
In particular, the VHE gamma-rays are expected to be observed once a $\sim 3$ years by CTAO, assuming the 10\% of its duty cycle and considering that $\sim$30 \% of GRBs have flares.
Additionally, Figure \ref{fig:r-Gamma} indicates where in the $r_{\rm diss}$-$\Gamma$ plane the UV to VHE gamma rays can be detected by Swift/UVOT and CTAO.
According to the  Figure \ref{fig:r-Gamma}, the detection of UV suggests $r_{\rm diss}>10^{13}$ cm, while the VHE gamma-ray detection does $\Gamma>100$ for our fiducial parameter sets.
Note that high value of magnetic field, $\xi_B$, and power-low slope, $p$, reduce the parameter range over which VHE gamma rays can be detected, as discussed in the Appendix. 
The actual detection and non-detection rates of multi-wavelength emission are important for constraining the uncertain yet essential parameters $r_{\rm diss}$ and $\Gamma$, which are key to understanding the physics of relativistic jets.
Future surveys of flares by Swift, SVOM, and CTAO may reveal these values.

\begin{acknowledgments} 
The authors thank Yuri Sato for meaningful discussions. 
This work is supported by Graduate Program on Physics for the Universe (GP-PU), and JSPS KAKENHI Nos. 25KJ0587 (R.M.) 22K14028, 21H04487, and 23H04899 (S.S.K.). 
S.S.K. acknowledges the support by the Tohoku Initiative for Fostering Global Researchers for Interdisciplinary Sciences (TI-FRIS) of MEXT's Strategic Professional Development Program for Young Researchers. 
The work of K.M. was supported by the NSF Grants No. AST-2108466, No. AST-2108467, and No. 2308021, and KAKENHI No. 20H05852. 
B.T.Z. is supported in China by National Key R\&D program of China under the grant 2024YFA1611402.
\end{acknowledgments}

\appendix
The parameters $\xi_B$, $p$, and $p_{\rm e,min}$ used in the main text are uncertain from both theoretical and observational perspectives.
To address their dependence on these parameters, Appendixes \ref{app:magjet}, \ref{app:p25}, and \ref{app:pmin20} discuss cases with different values of these parameters.

\section{Magnetized jet case}
\label{app:magjet}
In the main text, we discuss the dissipation of an unmagnetized jet in the limit $\xi_B < 1$, but it is also possible that magnetic field energy dominates the jet prior to dissipation.
Such a magnetized jet may dissipate its energy into electrons as it approaches equipartition.
To account for this limit, we adopt $\xi_B = 1$.

\begin{figure*}\hspace{-1cm}
\centering
\includegraphics[keepaspectratio, scale=0.7]{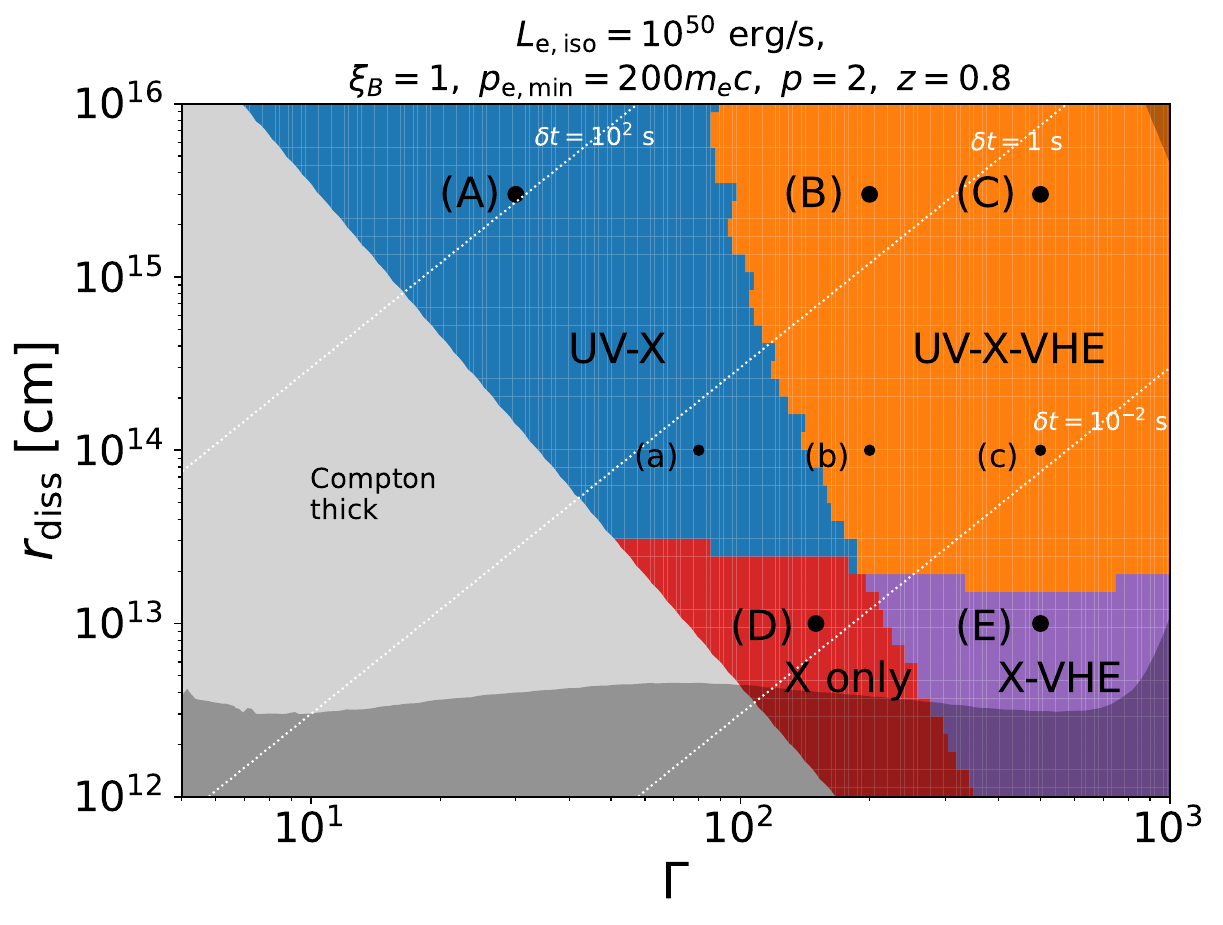}
\caption{Same as Figure \ref{fig:r-Gamma} but for the magnetized jet.}
\label{fig:r-Gamma_mag}
\end{figure*}

Figure \ref{fig:r-Gamma_mag} shows the detectability in the $r_{\rm diss}$-$ \Gamma$ plane for the magnetized jet case.
For these cases, the synchrotron burn-off limit around $\sim 8 ~\Gamma_2[(1+z)/2]^{-1}$ GeV become significant.
This produces an exponential cutoff above that energy.
This cutoff prevents us from detecting VHE gamma rays by CTAO for $\Gamma \sim 50$-$200$, which is the most significant difference comparing to the case with $\xi_B=10^{-1}$.

\section{soft electron injection case}
\label{app:p25}
\begin{figure*}\hspace{-0.5cm}
    \begin{tabular}{cc}
      \begin{minipage}[t]{0.5\hsize}
        \centering
        \includegraphics[keepaspectratio, scale=0.52]{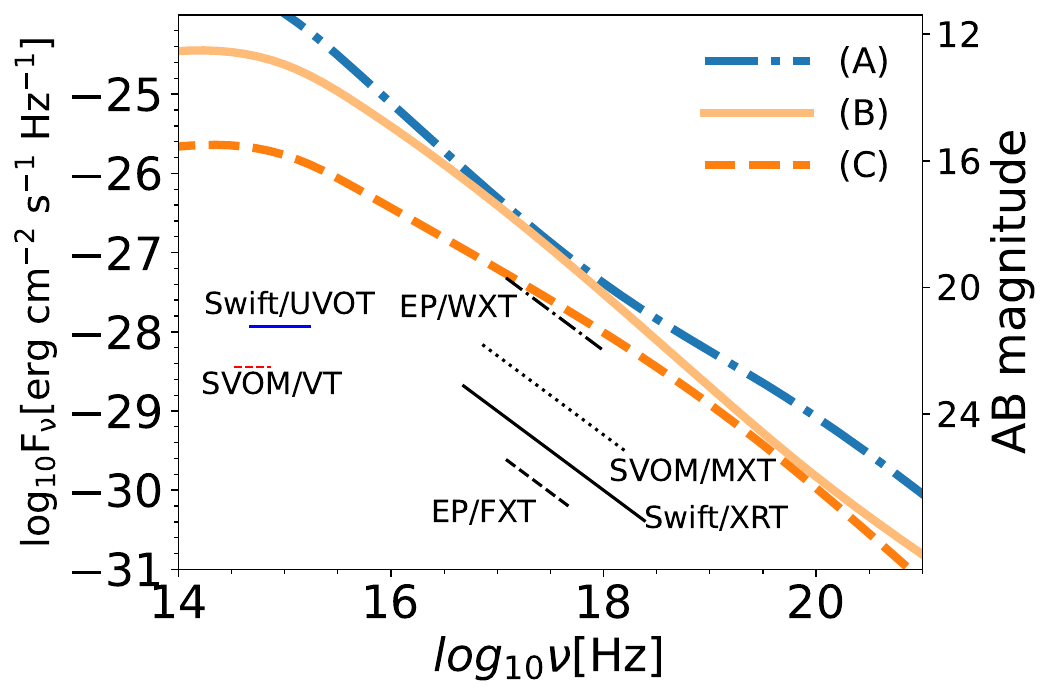}
      \end{minipage}
      
      \begin{minipage}[t]{0.5\hsize}
        \centering
        \includegraphics[keepaspectratio, scale=0.52]{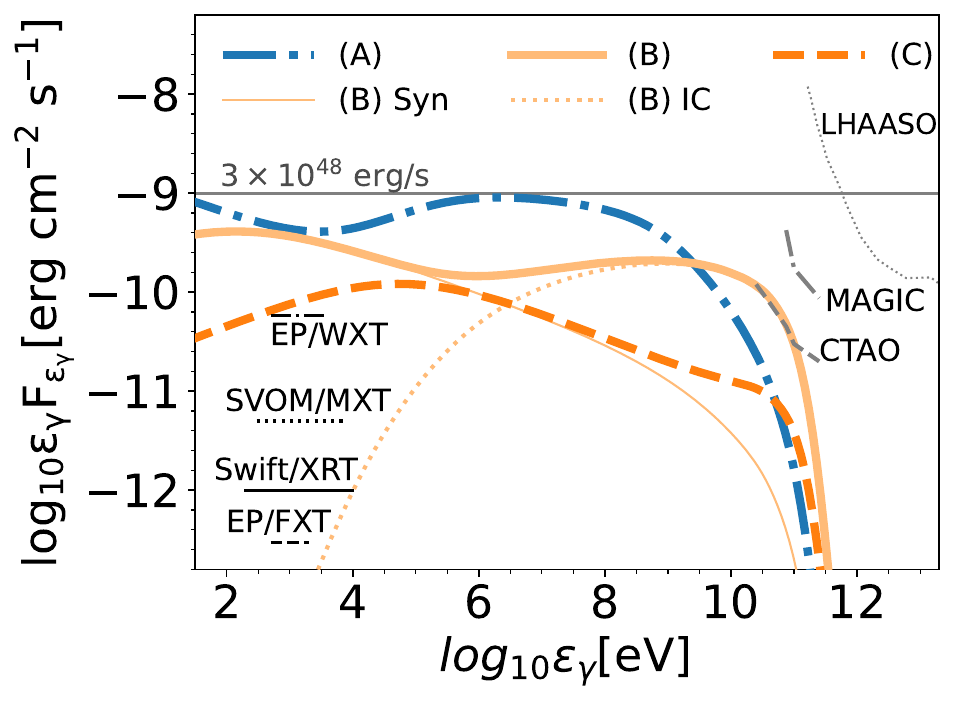}
      \end{minipage}
    \end{tabular}

    \begin{tabular}{cc}\hspace{-0.5cm}
      \begin{minipage}[t]{0.5\hsize}
        \centering
        \includegraphics[keepaspectratio, scale=0.52]{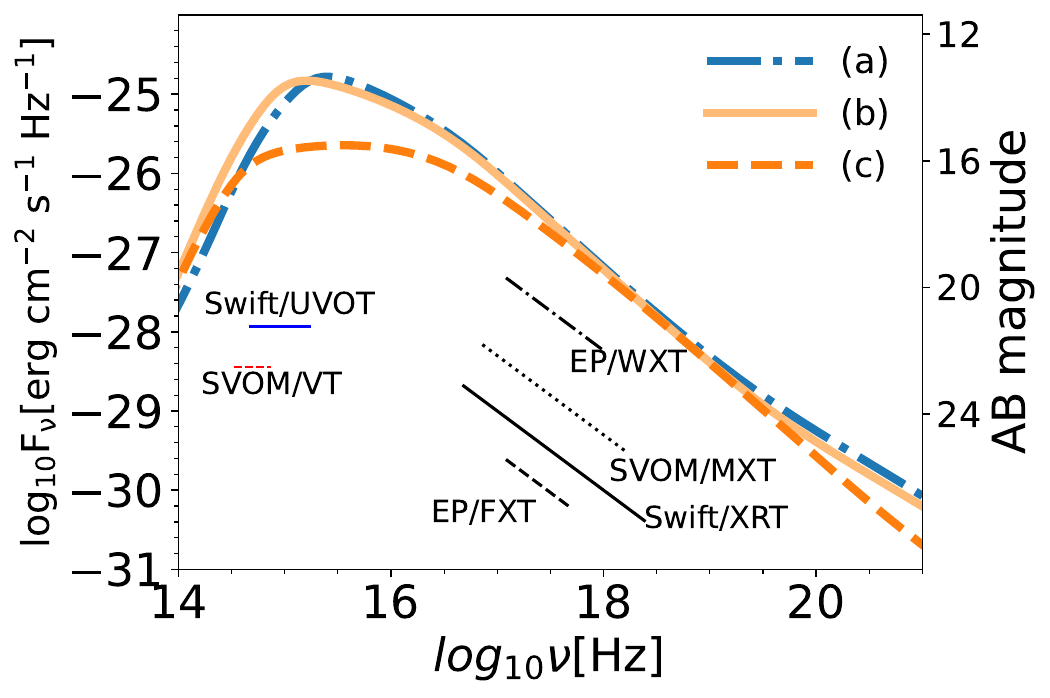}
      \end{minipage}
      
      \begin{minipage}[t]{0.5\hsize}
        \centering
        \includegraphics[keepaspectratio, scale=0.52]{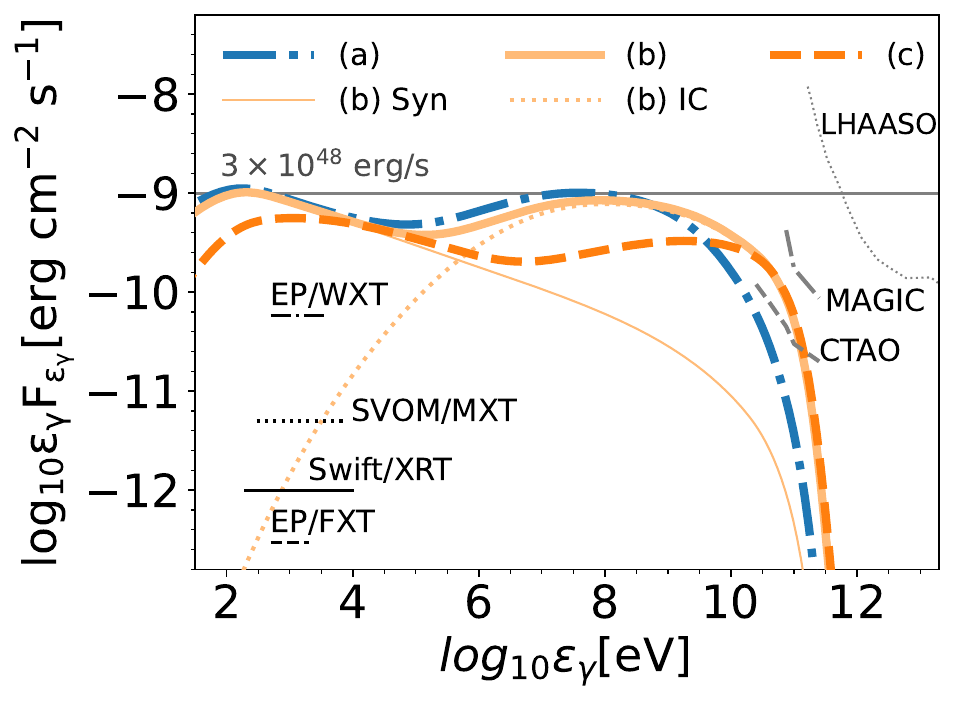}
      \end{minipage}
    \end{tabular}
    
    \begin{tabular}{cc}\hspace{-0.5cm}
      \begin{minipage}[t]{0.5\hsize}
        \centering
        \includegraphics[keepaspectratio, scale=0.52]{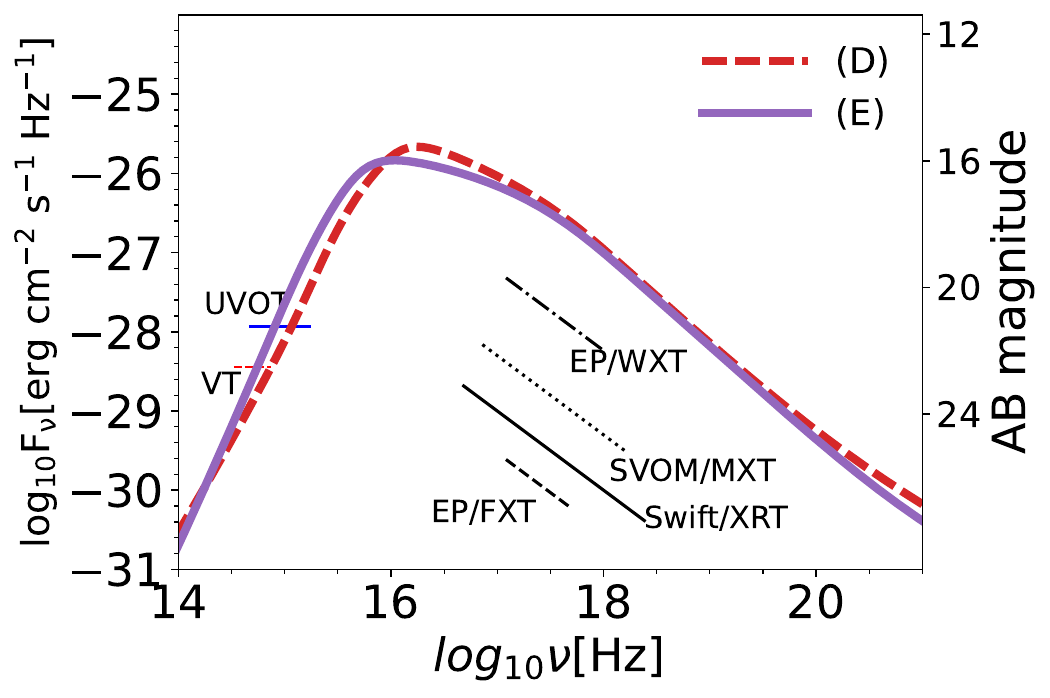}
      \end{minipage}
      
      \begin{minipage}[t]{0.5\hsize}
        \centering
        \includegraphics[keepaspectratio, scale=0.52]{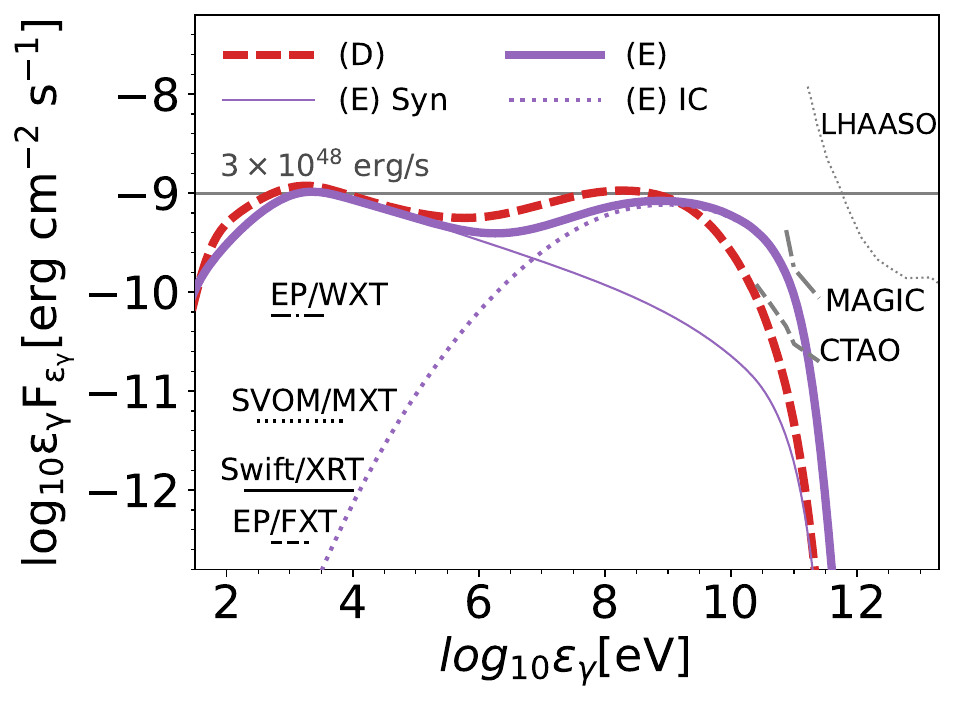}
      \end{minipage}
      
    \end{tabular}
    
    \caption{Same as Figure \ref{fig:spectra} but for the soft electron injection case ($p=2.5$).}.
    \label{fig:spectra_p25}
  \end{figure*}

In the main text, we assume an electron injection with $p = 2$, which results in hard spectra for both electrons and photons.
However, this value of $p$ is theoretically and observationally uncertain.
Here, we examine cases with $p = 2.5$ as a representative soft injection limit for XFs and EEs.
Figure \ref{fig:spectra_p25} shows the spectra for $p = 2.5$ with the same parameters as in Tables \ref{tab:param_a2f} and \ref{tab:param}.
A two-hump structure is visible in all spectra except for the case (C), produced by the primary synchrotron and its self-inverse Compton (SSC) components.
The SSC component is responsible for the detectability of VHE gamma rays.

Overall, the detectability is similar to that for $p = 2$, but the spectra for the case (C) do not reach the CTAO sensitivity.
This is because the case is in the slow-cooling regime, where the radiation efficiency is lower than 100\%.
The SSC component is suppressed as the primary synchrotron emissions, or seed photons, decrease.

\begin{figure*}\hspace{-1cm}
\centering
\includegraphics[keepaspectratio, scale=0.7]{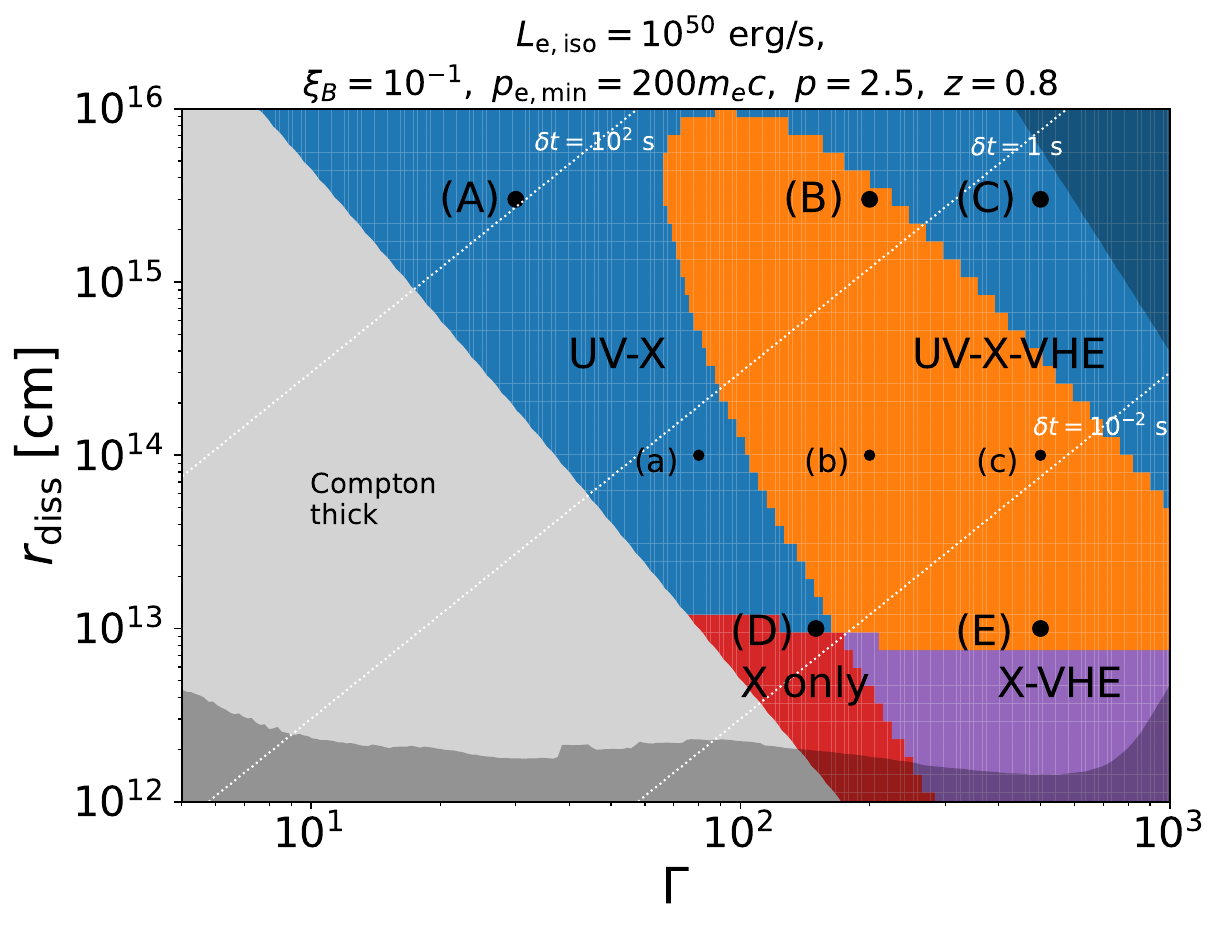}

\caption{Same as Figure \ref{fig:r-Gamma} but for $p=2.5$.}
\label{fig:r-Gamma_p25}
\end{figure*}

\begin{figure*}\hspace{-1cm}
\centering
\includegraphics[keepaspectratio, scale=0.7]{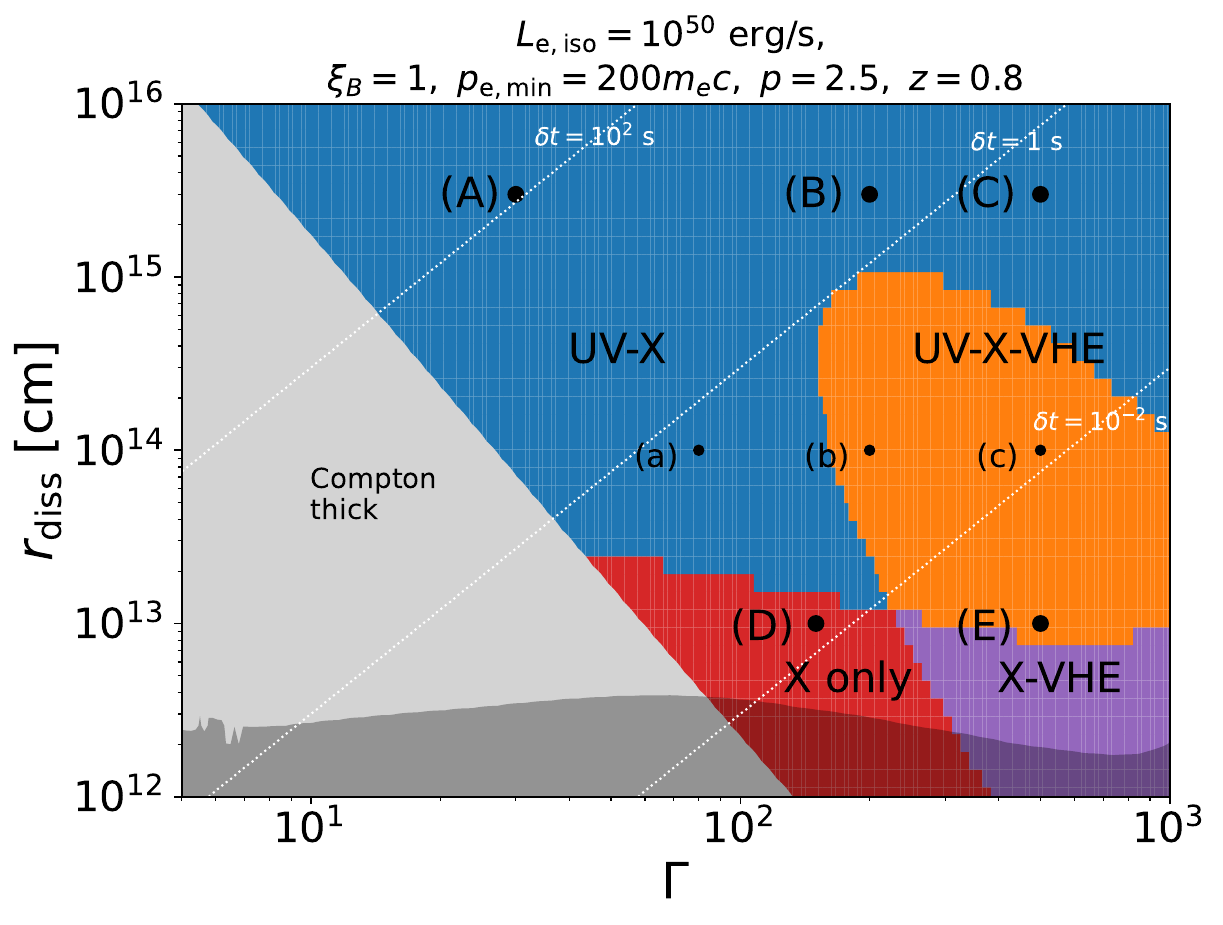}

\caption{Same as Figure \ref{fig:r-Gamma} but for $p=2.5$ and $\xi_B =10$.}
\label{fig:r-Gamma_magp25}
\end{figure*}

Figure \ref{fig:r-Gamma_p25} shows the detectability in the $r_{\rm diss}$-$\Gamma$ plane for the soft injection ($p=2.5$).
The detectability of VHE gamma rays are modified for around the case (C), high $\Gamma>300$-$500$ and large $r_{\rm diss}>10^{14}$-$10^{16}$ cm.
Figure \ref{fig:r-Gamma_magp25} shows the detectability in the $r_{\rm diss}$-$\Gamma$ plane for soft injection and magnetized jet case ($p=2.5$ and $\xi_B =10$).
Since the inverse-Compton emission is inefficient for $\xi_B=1$, the parameter space where the VHE gamma rays can be detected become narrow.
These plots make the complexity for constraining the $r_{\rm diss}$-$ \Gamma$ plane.
We need to clarify the value of $p$ from observed spectra when we constrain the $r_{\rm diss}$-$\Gamma$.

\section{Low injection energy case}
\label{app:pmin20}
In the main text, we assume a minimum momentum of the injected electrons, $p_{\rm e,min} = 200m_ec$.
However, this value is uncertain from both theoretical and observational perspectives.
To account for this uncertainty and to demonstrate that this parameter has only a minor impact on our results, this section presents the multi-wavelength spectra and detectability for $p_{\rm e,min} = 20$, which may arise from shock acceleration at the dissipation region.

\begin{figure*}\hspace{-0.5cm}
    \begin{tabular}{cc}
      \begin{minipage}[t]{0.5\hsize}
        \centering
        \includegraphics[keepaspectratio, scale=0.52]{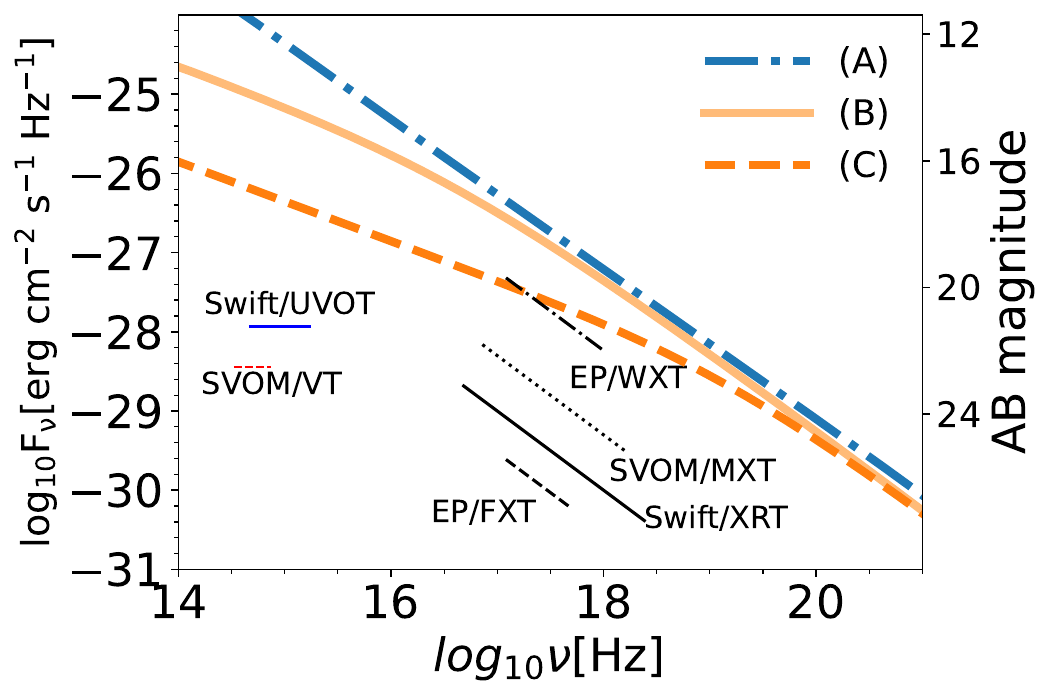}
      \end{minipage}
      
      \begin{minipage}[t]{0.5\hsize}
        \centering
        \includegraphics[keepaspectratio, scale=0.52]{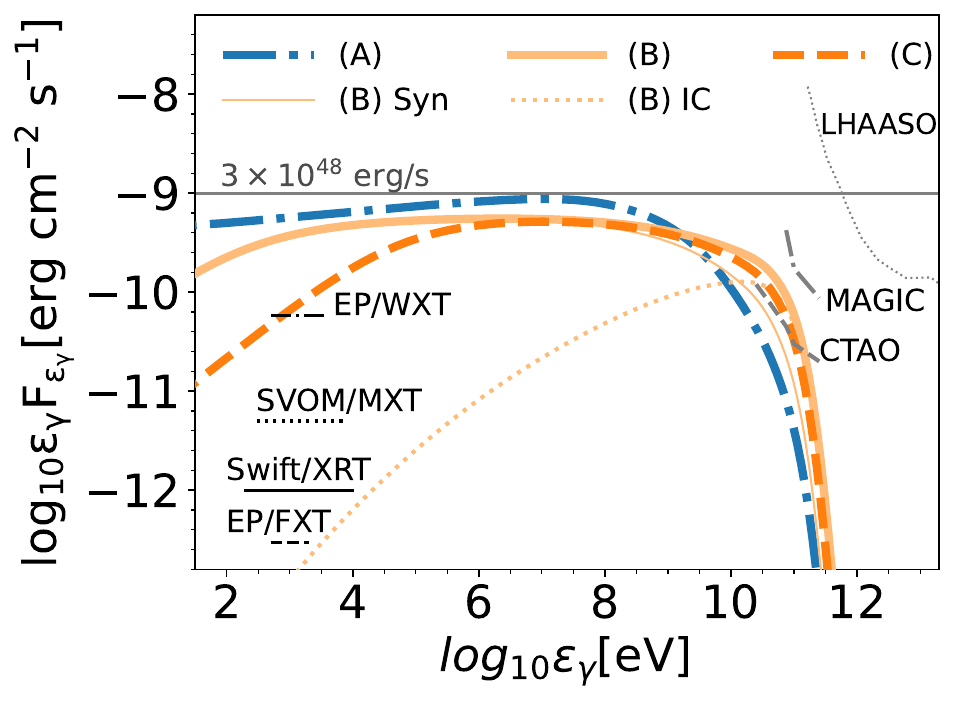}
      \end{minipage}
    \end{tabular}

    \begin{tabular}{cc}\hspace{-0.5cm}
      \begin{minipage}[t]{0.5\hsize}
        \centering
        \includegraphics[keepaspectratio, scale=0.52]{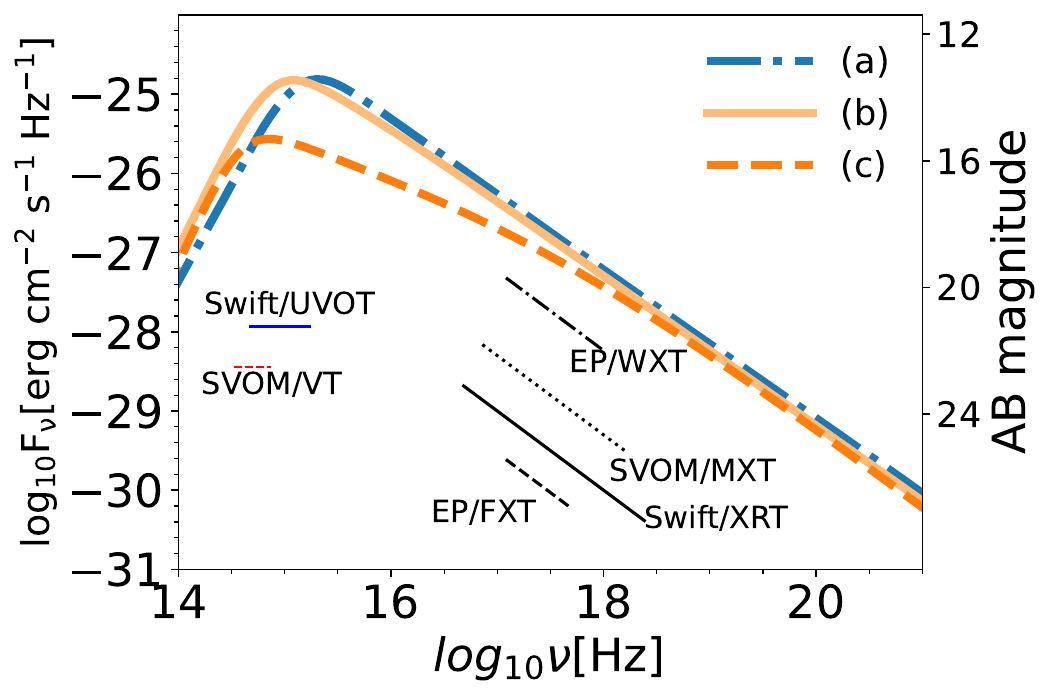}
      \end{minipage}
      
      \begin{minipage}[t]{0.5\hsize}
        \centering
        \includegraphics[keepaspectratio, scale=0.52]{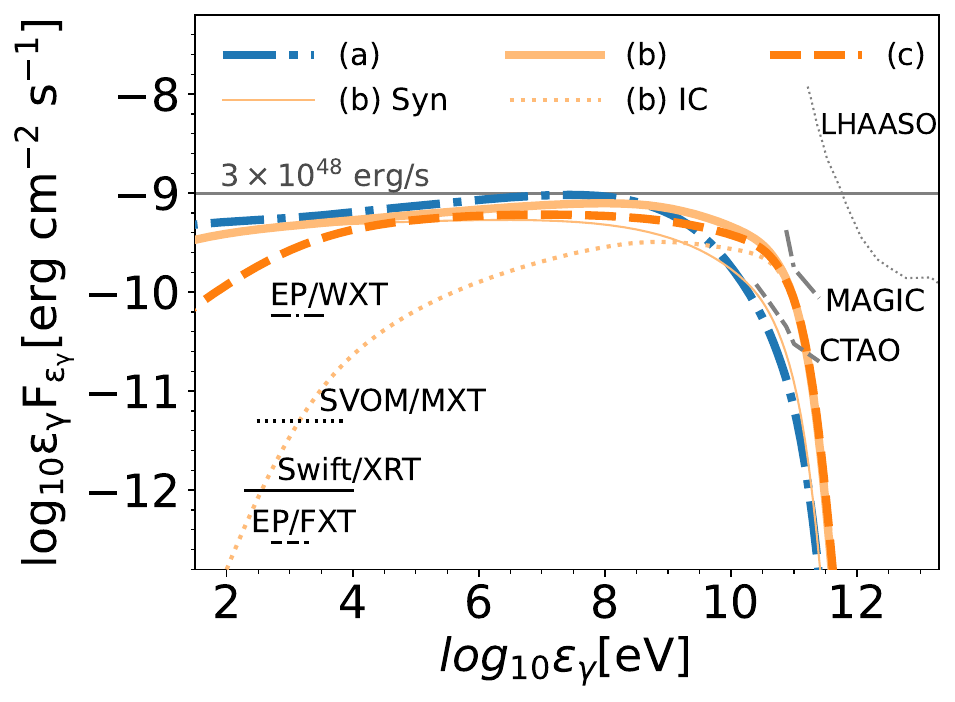}
      \end{minipage}
    \end{tabular}
    
    \begin{tabular}{cc}\hspace{-0.5cm}
      \begin{minipage}[t]{0.5\hsize}
        \centering
        \includegraphics[keepaspectratio, scale=0.52]{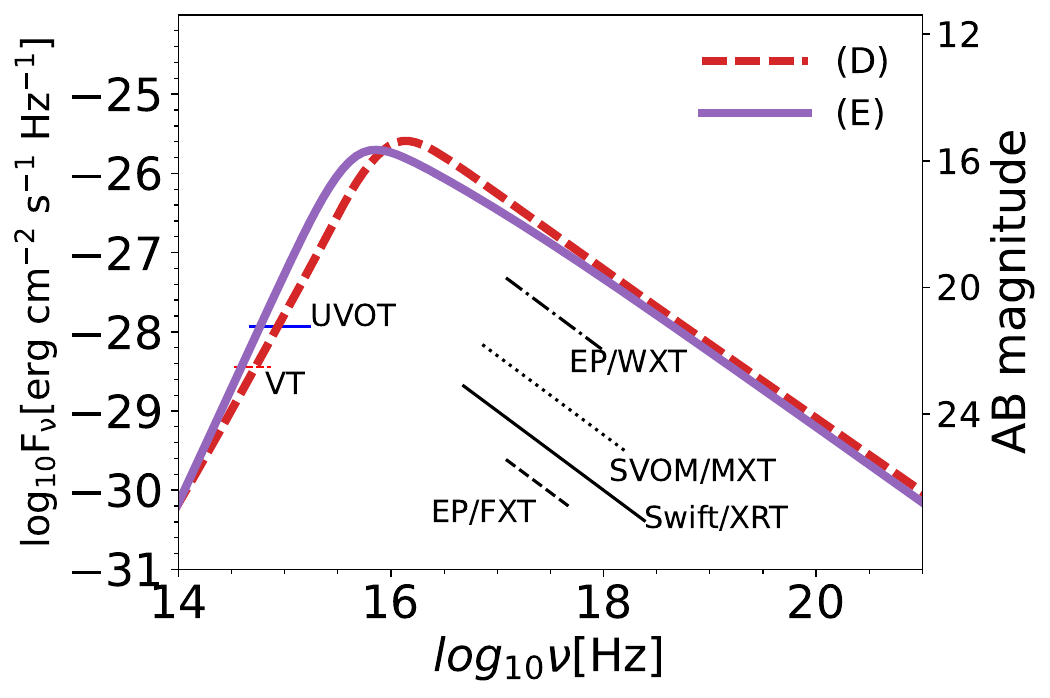}
      \end{minipage}
      
      \begin{minipage}[t]{0.5\hsize}
        \centering
        \includegraphics[keepaspectratio, scale=0.52]{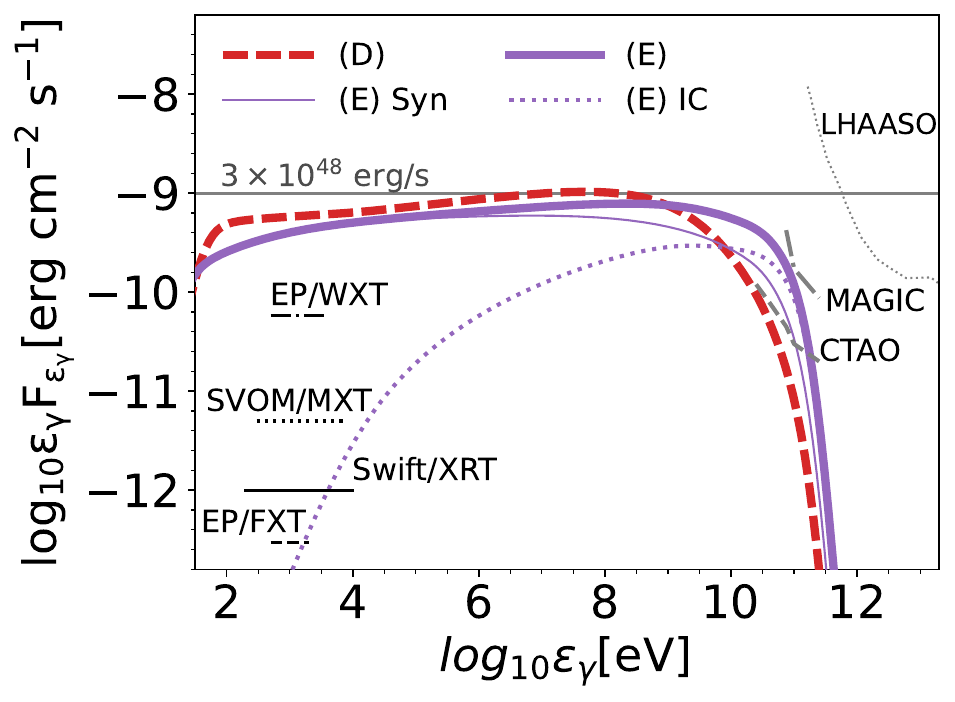}
      \end{minipage}
      
    \end{tabular}
    
    \caption{Same as Figure \ref{fig:spectra} but for the low injection energy case ($p_{\rm e,min}=20m_ec$).}.
    \label{fig:spectra_pmin20}
  \end{figure*}

Figure \ref{fig:spectra_pmin20} shows the spectra calculated with $p_{\rm e,min} = 20$.
These spectra differ slightly from those in Figure \ref{fig:spectra}, mainly in the spectral break energy at $100~{\rm eV} - 100~{\rm keV}$, where the spectra deviate from the relations $\varepsilon_\gamma F_{\varepsilon_\gamma}\propto \varepsilon_\gamma^0$ or $F_\nu \propto \nu^{-1}$.
Since cases (C) and (c) are in the slow-cooling regime, the break energy does not depend on the injection energy and thus is independent of $p_{\rm e,min}$.
On the other hand, for cases (A), (B), (a), (b), (E), and (F), which are in the fast-cooling regime, the break energy is directly determined by $p_{\rm e,min}$, and a lower $p_{\rm e,min}$ leads to a lower break energy.
Although the lower break energy slightly enhances the UV emission, this parameter does not significantly affect the detectability in the X-ray band.

\begin{figure*}\hspace{-1cm}
\centering
\includegraphics[keepaspectratio, scale=0.7]{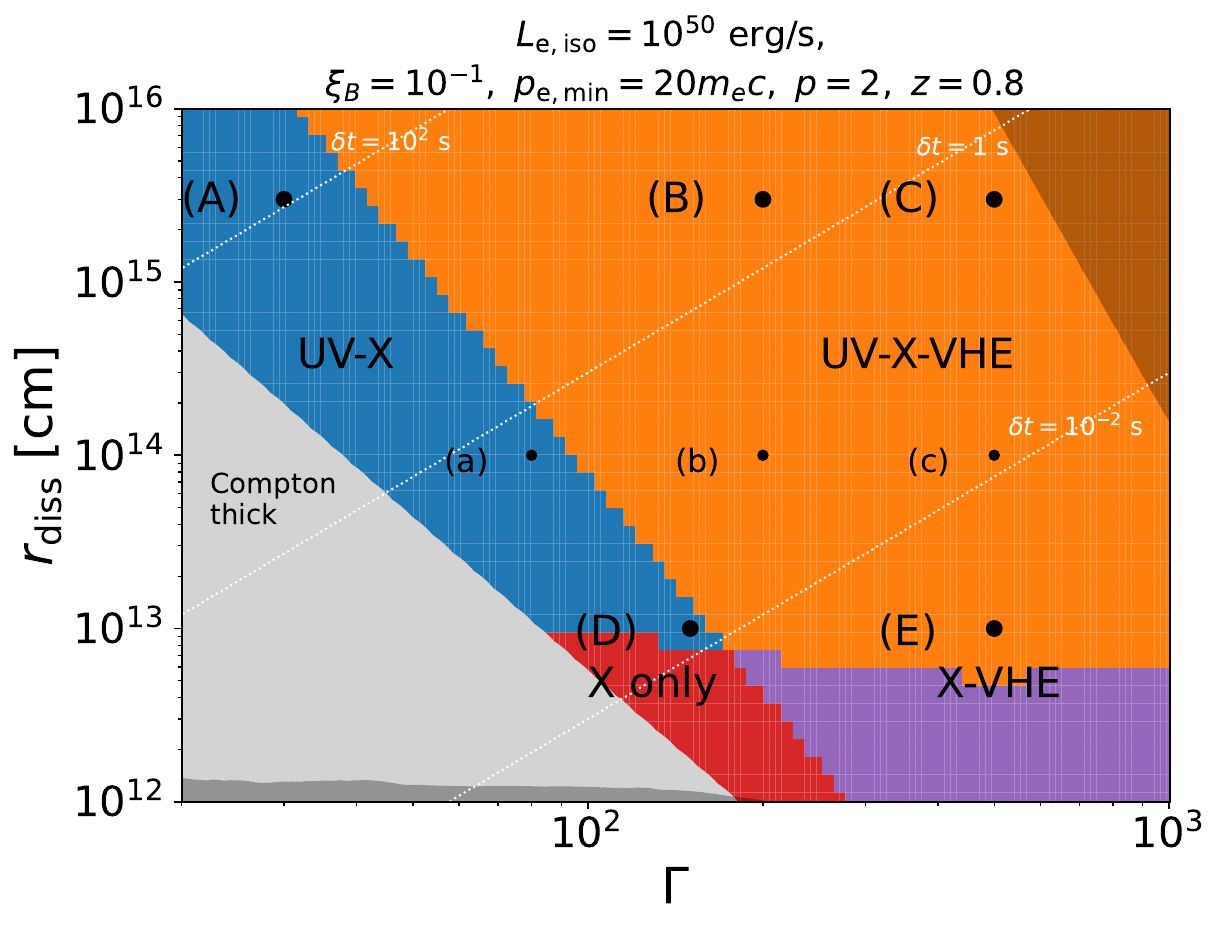}

\caption{Same as Figure \ref{fig:r-Gamma} but for $p_{\rm e,min}=20m_ec$.}
\label{fig:r-Gamma_pmin20}
\end{figure*}

The detectability in the UV and VHE gamma-ray bands is summarized in Figure \ref{fig:r-Gamma_pmin20}.
The detectable range in the UV (blue and orange regions) becomes slightly broader, and the UV emission becomes detectable for cases (D) and (E).
In contrast, the detectability of VHE gamma rays is less sensitive to the value of $p_{\rm e,min}$.

\bibliography{sample701}{}
\bibliographystyle{aasjournalv7}



\end{document}